\documentclass{article}

\usepackage[english]{babel}

\usepackage[letterpaper,top=2cm,bottom=2cm,left=3cm,right=3cm,marginparwidth=1.75cm]{geometry}

\usepackage{amsmath,amssymb,mathrsfs}
\usepackage{graphicx}
\usepackage{authblk}
\usepackage{empheq}
\usepackage{subcaption,ragged2e}
\usepackage[colorlinks=true, allcolors=blue]{hyperref}

\author[1]{Elena Vernazza\thanks{Now at CERN, European Organization for Nuclear Research, Geneva, Switzerland}}
\author[2]{Jona Motta}
\author[1]{Léa-Maria Rabour}
\author[1]{Shamik Ghosh}
\author[1]{Emilia Becheva}
\author[1]{Maria Frau}
\author[1]{Théophile Le Clerc}
\author[1]{Frédéric Magniette}
\author[1]{Jean-Baptiste Sauvan}
\author[1]{Olivier Davignon}

\affil[1]{Laboratoire Leprince-Ringuet, CNRS/IN2P3, \'Ecole polytechnique, Institut Polytechnique de Paris, Palaiseau, France}
\affil[2]{Universität Zürich, Zurich, Switzerland}

\title{The Calibr-A-Ton: a novel method\\for calorimeter energy calibration}

\begin{document}
\maketitle

\begin{abstract}
The energy calibration of calorimeters at collider experiments, such as the ones at the CERN Large Hadron Collider, is crucial for achieving the experiments physics objectives. Standard calibration approaches have limitations that become more pronounced as detector granularity increases. In this paper we propose a novel calibration procedure to simultaneously calibrate individual detector cells belonging to a particle shower by targeting a well-controlled energy reference. The method bypasses some of the difficulties that exist in more standard approaches, and it is implemented using differentiable programming. Simulated energy deposits in the electromagnetic section of a high-granularity calorimeter are used to study the method and demonstrate its performance. It is shown that the method is able to correct for biases in the energy response.
\end{abstract}

\section{Introduction}

The energy calibration of a calorimeter consists in methods that allow to precisely estimate the energy of particles that deposit energy within its volume. An imprecise calorimeter calibration can induce biases and uncertainties on the measurement of particles energies and hence, can significantly reduce the discovery potential of a physics experiment.\\

Since calorimeters are segmented into detection sub-components, commonly called ``cells'', the first step of an energy calibration consists in finding, for each cell, the factor that allows a precise correspondence between the signal given by that cell and the genuine energy deposit that occurred in that cell. These factors are typically called ``calibration constants''. The main goal of a calibration method is to perform a regression of these constants, as illustrated by the relation:
\begin{equation}
    C_{\mathrm{pos},S_\mathrm{pos}}\times S_\mathrm{pos} \rightarrow E_\mathrm{ref},
    \label{cell}
\end{equation}
where ``pos'' refers to the $x,y$ and $z$ coordinates of the cell, $S_\mathrm{pos}$ the signal given by that cell, $C_{\mathrm{pos},S_\mathrm{pos}}$ the (discrete) calibration constants for that cell with signal $S_\mathrm{pos}$, and $E_\mathrm{ref}$ is the reference energy deposited in that cell (for example, the true/generator energy). The number of $C_{\mathrm{pos},S_{\mathrm{pos}}}$ depends on the binning chosen for the pos and $S$ variables. The goal of the calibration is to find the set of calibration constants that minimize the discrepancy between the calibrated energy (left side of Eq. \ref{cell}) and the reference (right side of Eq. \ref{cell}).\\

Taking the CMS electromagnetic calorimeter (ECAL) as an example \cite{CMS:2024ppo}, the calibration constants are derived by laser monitoring, \emph{i.e.} using a laser as a standard source of light to induce a response in individual ECAL cells (lead-tungsten scintillating crystals), and by physics events such as photons from $\pi^0$ or $\eta$ meson decays, or electrons from $W$ or $Z$-boson decays. When using a laser as a proxy of the light produced by an energy deposit, the reference object corresponds, by construction, to a single cell. However, when using physics signals, such one-to-one correspondence does not necessarily exist, since particles tend to deposit their energies in multiple calorimeter cells (typically $\mathcal{O}(10)$ cells for the multiple-GeV electrons used in the CMS ECAL calibration). For that reason, the usual calibration procedures focus on calibrating the cell that collects the most energy. Such compromises become extremely challenging when high-granularity sampling calorimeters with millions of channels are considered, for instance the High-Granularity Calorimeter (HGCAL) of the CMS experiment \cite{HGCAL}. For example, it is unimaginable to design a system that would point a laser in every cell of the detector (\emph{e.g.} HGCAL contains $\sim$6 million cells). In addition, in HGCAL, multiple-GeV electrons deposit their energies in $\mathcal{O}(500)$ cells with very large fluctuations in the location and energy distributions and, in that context, focusing the calibration on the cell that collects the most energy is very statistically inefficient: for every particle there is a large amount of information to be learned from the deposits in the $\mathcal{O}(500)$ other cells.\\

The Calibr-A-Ton method, described in section \ref{sec:desc}, aims at overcoming some of those limitations. To evaluate the performance of that method, a high granularity sampling calorimeter design, that consists of a hexagonal electromagnetic section -- aimed at the detection of electrons and photons -- and two hexagonal hadronic sections placed in succession -- aimed at the detection of hadrons -- was developed within the \texttt{GEANT4} software \cite{GEANT4}. This simple calorimeter design is described shortly in section \ref{sec:calo}. Section \ref{sec:biases} describes the introduction of known biases on the cell energy response and their effects on the energy resolution of the simulated calorimeter. In section \ref{performance}, the performance of Calibr-A-Ton is discussed in terms of its capability to correct for the known biases introduced. For illustration, the Calibr-A-Ton procedure is compared to a method where the calibration constants are derived using a regression of lower dimensionality. A conclusion is given in section \ref{conclusion}.

\section{Description of the Calibr-A-Ton method}
\label{sec:desc}
Relation \ref{cell} can be generalized to all cells belonging to the particle shower:
\begin{equation}
    \sum_{i=1}^{n_\mathrm{cells}} C_\mathrm{pos^i,S^i_\mathrm{pos}}\times S^i_\mathrm{pos} \rightarrow E_\mathrm{ref},
    \label{cells}
\end{equation}
where $i$ are labels for the cells\footnote{The number of cells that the method considers as belonging to the particle shower is a tunable parameter: considering a single cell may not capture the whole energy of the shower, while considering too many might lead to integrating more deposits from other particles or more electronics noise.} in which a non-vanishing signal, \emph{i.e.} an energy deposit, is detected. Using this method, every shower allows to constrain multiple $C_\mathrm{pos,S_\mathrm{pos}}$ values and therefore less statistics is necessary to obtain an arbitrary precision on the calibration constants. In addition, compared to the usual methods, there is a less stringent need to tightly select showers with given topological properties (for example selecting deposits that occur in a low number of cells) -- many more showers (in particular high cell multiplicity showers) can be used, further increasing the statistical power of the calibration procedure.\\

For a set of $N_\mathrm{showers}$ distinct calorimeter showers, called the ``training'' sample, the set of  $C_\mathrm{pos,S_\mathrm{pos}}$ parameters can be estimated by minimizing a loss function, $\mathscr{L}$, which can be of the form:
\begin{equation}
    \mathscr{L}\;\hat{=}\;\sum_{j=0}^{N_\mathrm{showers}} \frac{\left | \sum_{i} C_\mathrm{pos^{i,j},S^{i,j}_{pos}}\times S^{i,j}_\mathrm{pos} - E^j_\mathrm{ref}\right |}{E^j_\mathrm{ref}},
    \label{eq:loss}
\end{equation}
where $\mathrm{pos^{i,j}}$ are the spatial coordinate labels for the cell $i$ belonging to the shower $j$; $S^{i,j}_\mathrm{pos}$ is its signal. 

The minimization is performed by employing gradient descent optimization using the Adam \cite{kingma2014adam} optimizer which adaptively adjusts the learning rate per parameter resulting in efficient convergence. The training sample is divided in batches of equal number of showers and the calibration constants are updated at the end of every batch. The learning rate is chosen to depend on the batch number using \emph{cosine annealing}, a scheduling strategy which reduces the learning rate gradually following a cosine decay function to improve convergence stability.\\

The Calibr-A-Ton method had already been explored in a different context in the Ph.D. theses of Jona Motta \cite{Motta} and Elena Vernazza \cite{Vernazza}.\\

The performance of Calibr-A-Ton, in which the calibration constants are regressed in bins of the cells energies and 3-dimensional positions, is compared to a standard approach, frequently used in sampling calorimeters, where only a single calibration constant per layer is derived (called \emph{per-layer calibration}). In that method, the loss function of Eq. \ref{eq:loss} simplifies to:
\begin{equation}
    \mathscr{L}_{\ell}\;\hat{=}\;\sum_{j=0}^{N_\mathrm{showers}} \frac{\left | \sum_\ell C_{\ell}\times S^{j}_{\ell} - E^j_\mathrm{ref}\right |}{E^j_\mathrm{ref}},
    \label{eq:loss_layer}
\end{equation}
where $S^{j}_{\ell}$ is the sum of the cells energies in layer $\ell$ and $C_{\ell}$ the per-layer calibration constants.

\section{Calorimeter simulation and event simulation}
\label{sec:calo}

\subsection{Calorimeter concept}

\begin{figure}[h!]
\begin{center}
\includegraphics[height=4cm]{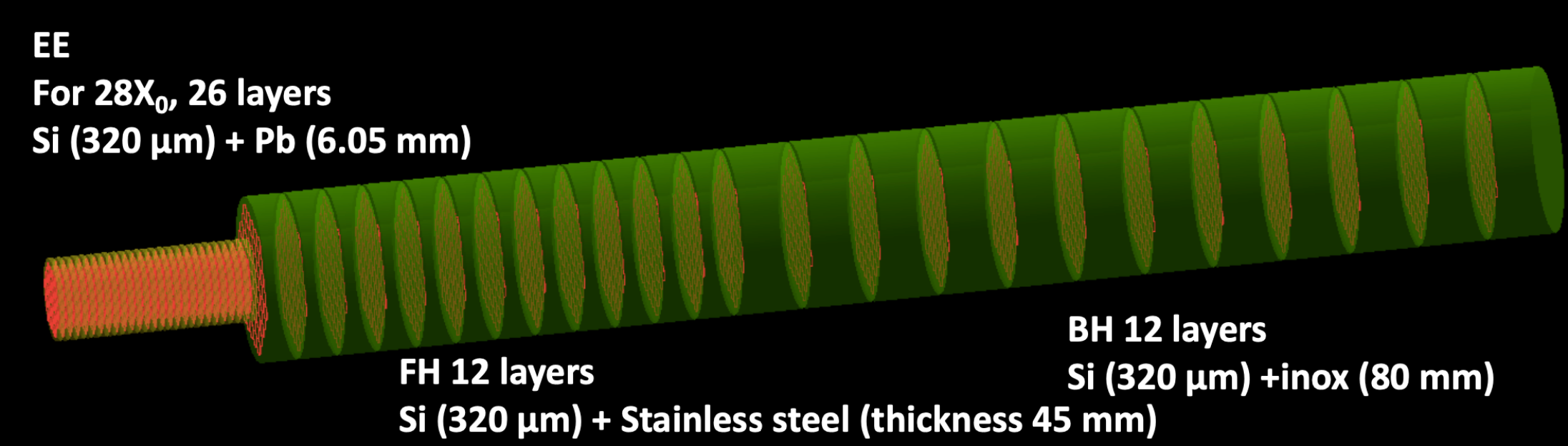}
\caption{Geometry of the \texttt{d2} calorimeter used in this paper. It features three hexagonal sections: EE, aimed at the detection of electrons and photons; and FH and BH, aimed at the detection of hadrons.}
\label{geometry1}
\end{center}
\end{figure}

\begin{figure}[h!]
\begin{center}
\includegraphics[height=7.5cm]{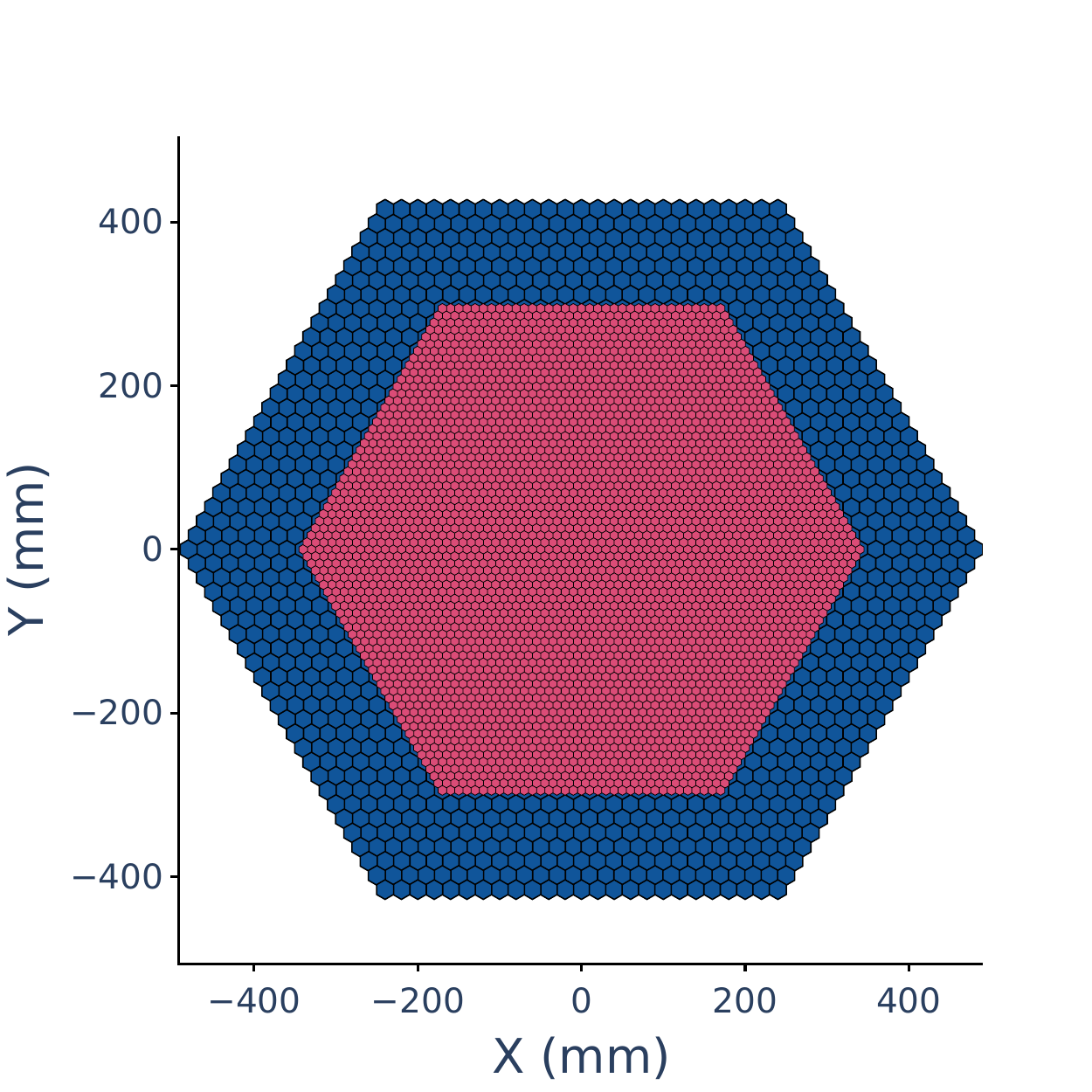}
\includegraphics[height=7.5cm]{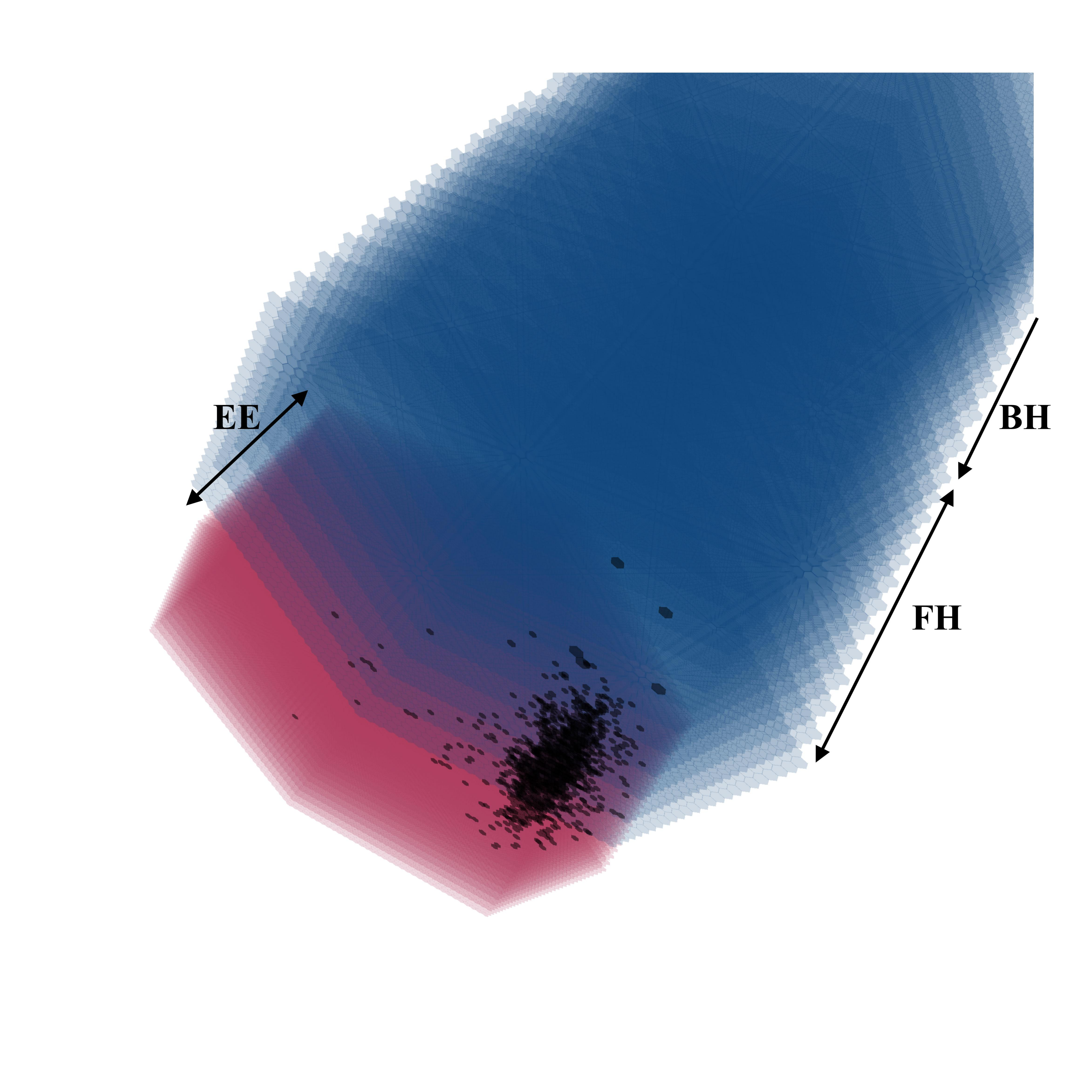}

\caption{(Left) Front view (along the $z$-axis) of the calorimeter. The cells of the EE section are visible in the middle of the figure, in red; they are hiding some of the bigger cells of the FH section, while some FH cells can still be seen towards the edges of the cylinder, in blue. (Right) Shower in the electromagnetic section, resulting from the interaction of an incoming electron with the detector. The black dots correspond to energy deposits in individual detector cells. The region with the highest density of such points corresponds to the area where the majority of the electron energy is deposited.}
\label{layers}
\end{center}
\end{figure}

The high-granularity sampling calorimeter design used in this paper, called \texttt{d2}, comes from the OGCID project \cite{OGCID} and is inspired by the CMS HGCAL. As illustrated in Figures \ref{geometry1} and \ref{layers} (left), the \texttt{d2} design features a fully hexagonal geometry, and includes:
\begin{itemize}
    \item An electromagnetic section (EE), aimed at the detection of electrons and photons, which consists of 26 successive layers of 320~$\mathrm{\mu m}$ of silicon (the active sensor elements) plus 6.05~$\mathrm{mm}$ of lead (the absorber), totalizing $28$ radiation lengths ($X_0$). Each layer features 3571 hexagonal silicon cells arranged concentrically in 35 rings (each EE cell can be inscribed in a 10~mm radius circle). The layers are 3~mm distant from one another and the gaps are filled by air.
    \item Two hadronic sections, one directly behind EE called FH, and one in succession called BH, aimed at the detection of hadrons. The FH (BH) consists of 12 (12) layers of 320~$\mathrm{\mu m}$ of silicon (the active sensor elements) plus 45~$\mathrm{mm}$ (80~mm) of stainless steel (the absorber). Each layer features 1801 hexagonal silicon cells, arranged concentrically in 25 rings (each FH/BH cell can be inscribed in a 20~mm radius circle). In both FH and BH, the layers are 3~mm distant from one another and the gaps are filled by air.
\end{itemize}
Figure \ref{layers} (right) illustrates the interaction of an electron in the EE of the calorimeter, that produces a high cell multiplicity shower.\\

The geometry described above, along with the selected materials for the different layers, is implemented in the \texttt{GEANT4} software. The 3D position of an EE cell is defined using 3 variables: the layer number $\ell$ (=1 for the front layer and =26 for the last layer, that is the one adjacent with the FH); the ring number $r$ (=1 for the single central cell in any given layer, and =35 for the outermost ring of cells in any given layer); and the angle $\theta$, which is the polar angle in the $(x,y)$ plane. For simplicity, $\theta$ is binned into $204$ labels (one for each cell in ring \#35) called $\theta_k$. For any given cell, the assigned $\theta_k$  corresponds to the direction closest to its center, such that a unique set of ($r,\ell,\theta_k$) values unequivocally define the position of a unique cell in the EE.

\subsection{Simulated events}

\begin{figure}[h!]
\begin{center}
\includegraphics[height=7cm]{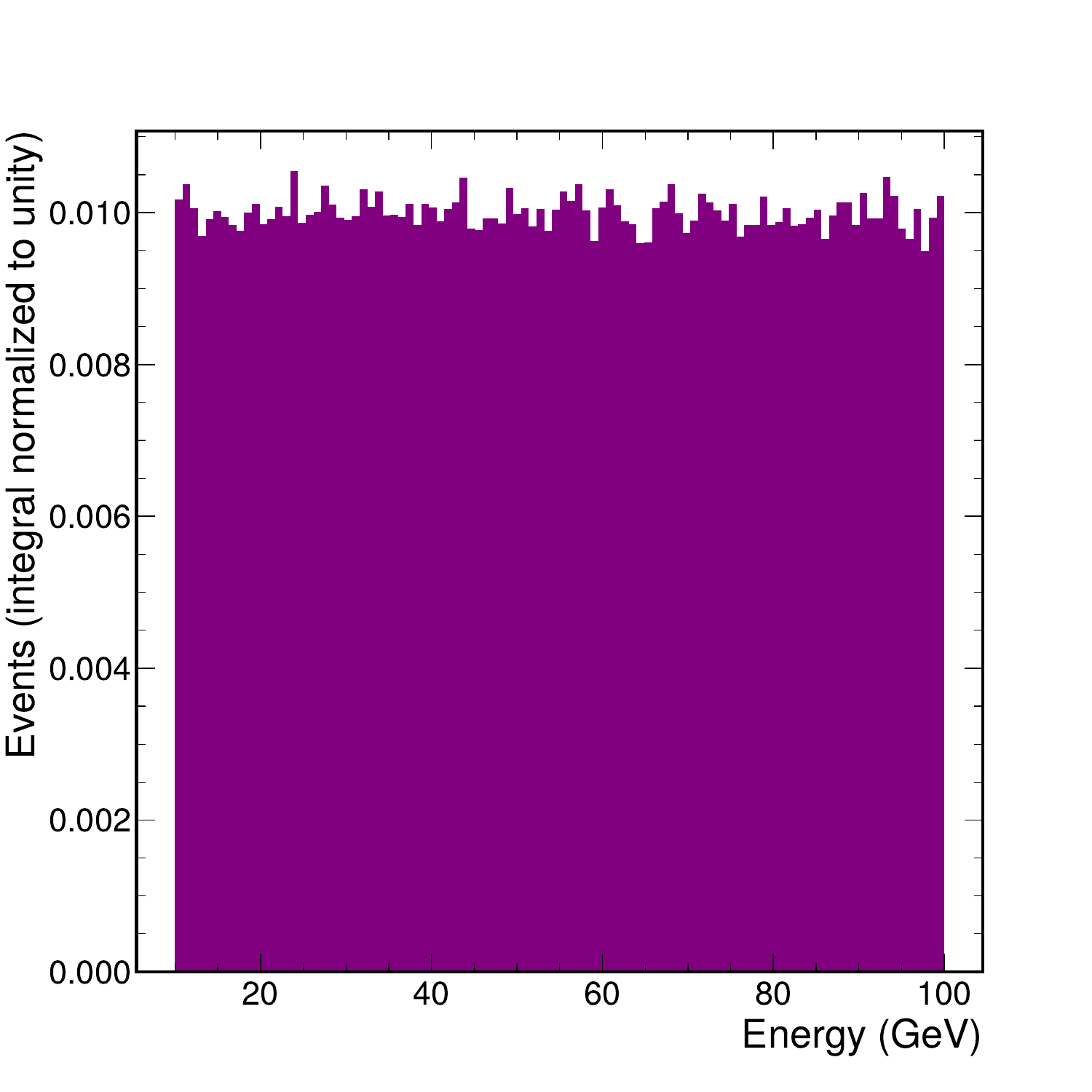}
\caption{Energy of the simulated electrons that are shot at the calorimeter used for the study. The spectrum is flat between 10 and 100 GeV.}
\label{egen}
\end{center}
\end{figure}

To assess the performance of the Calibr-A-Ton method, $5 \times 10^5$ electrons of known energies, uniformly distributed between 10 and 100 GeV, were simulated and shot perpendicularly towards the EE front surface. The energy spectrum of the simulated electrons is shown in Figure \ref{egen}. To ensure high enough statistics across the whole detector, the simulation is made such that about 10k electrons are shot along a trajectory that passes through each cell of the EE front surface. The signals from each silicon cell is taken as the raw energy deposit in the silicon, and is provided by \texttt{GEANT4}. Only cells with an energy deposit greater than $E_\mathrm{thr}=14.91$~keV are considered; the energy of the other ones is set to zero. The threshold chosen corresponds approximately to a tenth of the energy deposited by a minimum ionizing particle (MIP) in a 320~$\mathrm{\mu m}$-thick silicon cell\footnote{$E_\mathrm{thr}=10\%\times\mathrm{ t \times d_{Si} \times S}$, where $\mathrm{t=320\;\mu m}$ is the silicon sensor thickness, $d_\mathrm{Si}=2.33$~$\mathrm{g\cdot cm^{-3}}$ is the silicon density and $\mathrm{S}=2\;\mathrm{MeV\cdot cm^{2}\cdot g^{-1}}$ is the typical mean energy loss of a MIP particle.}. It is applied to reflect the typical dynamic range found in sampling calorimeters readout. Indeed, MIP are often used to assess the absolute energy scale of the detector, and must therefore be typically measured with a signal to noise ratio greater than five (for example, the electronics noise in HGCAL cells corresponds to roughly 20\% of a MIP energy deposit).

\subsection{Correction for the absorber}
\label{subsec:absorber_correction}

\begin{figure}[h!]
\begin{center}
\includegraphics[height=7cm]{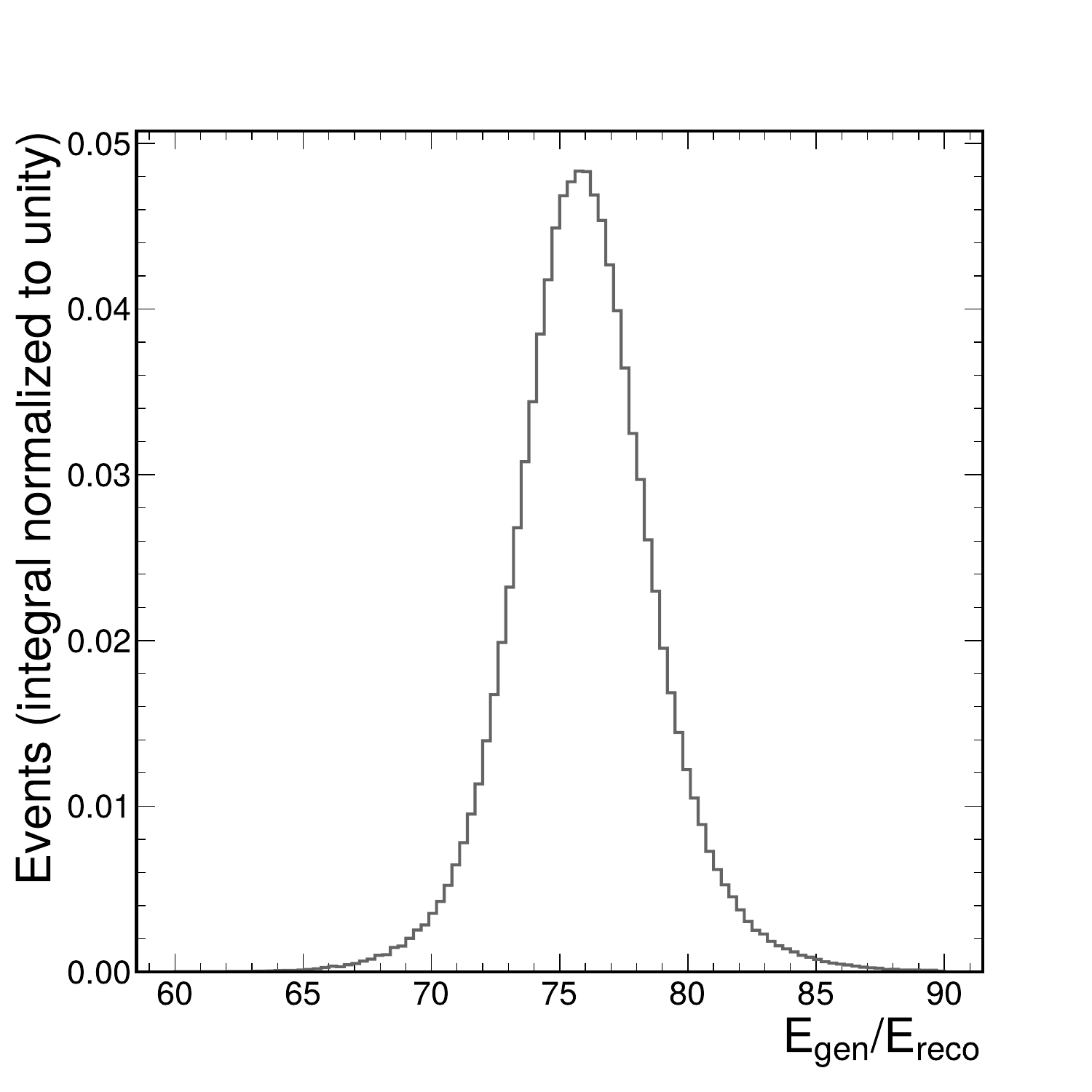}
\caption{Distribution of 1/response, as defined in Eq. \ref{eq:response}, for electron showers. The mean of that distribution is $\sim75.9$ and is used to correct for the presence of the absorber.}
\label{absorbercorrection}
\end{center}
\end{figure}

By design in sampling calorimeters, most of the electron energy is deposited within the layers of absorbers; the active volume, here silicon, only collects a small fraction of it. To account for that effect in the detector under study, a constant, called the \emph{absorber correction}, that is common to every cell in the EE, is computed from the mean of the $1/\mathrm{response}$ distribution. For a given shower, the response is defined as the following variable:
\begin{equation}
\label{eq:response}
    \mathrm{response}\;\hat{=}\;\frac{\sum_{i} C_{\ell^{i},r^{i},\theta_k^{i},S^{i}_{l,r,\theta_k}}\times S^{i}_{l,r,\theta_k}}{E_\mathrm{ref}}.
\end{equation}
The distribution of 1/response, when no calibration constants are applied ($C=1$), is shown in Figure \ref{absorbercorrection}. For convenience, the absorber correction, which is roughly equal to 75.9, is applied as a multiplicative factor to every cell signal, such that (on average and up to inter-calibration of the cells, fluctuations and noise effects) the energy of one generator particle is equal to the energy deposited in the calorimeter. In that context, after application of the absorber correction, one can expect the calibration constants to be very close to unity in the absence of any significant biases in the cell energy response. In the following, we assume that the absorber correction is always applied to all cells energies, and therefore it is omitted from equations.

\subsection{Binning}
\label{subsec:binning} 

\begin{figure}
     \centering
     \begin{subfigure}[b]{0.40\textwidth}
         \centering
         \includegraphics[width=\textwidth]{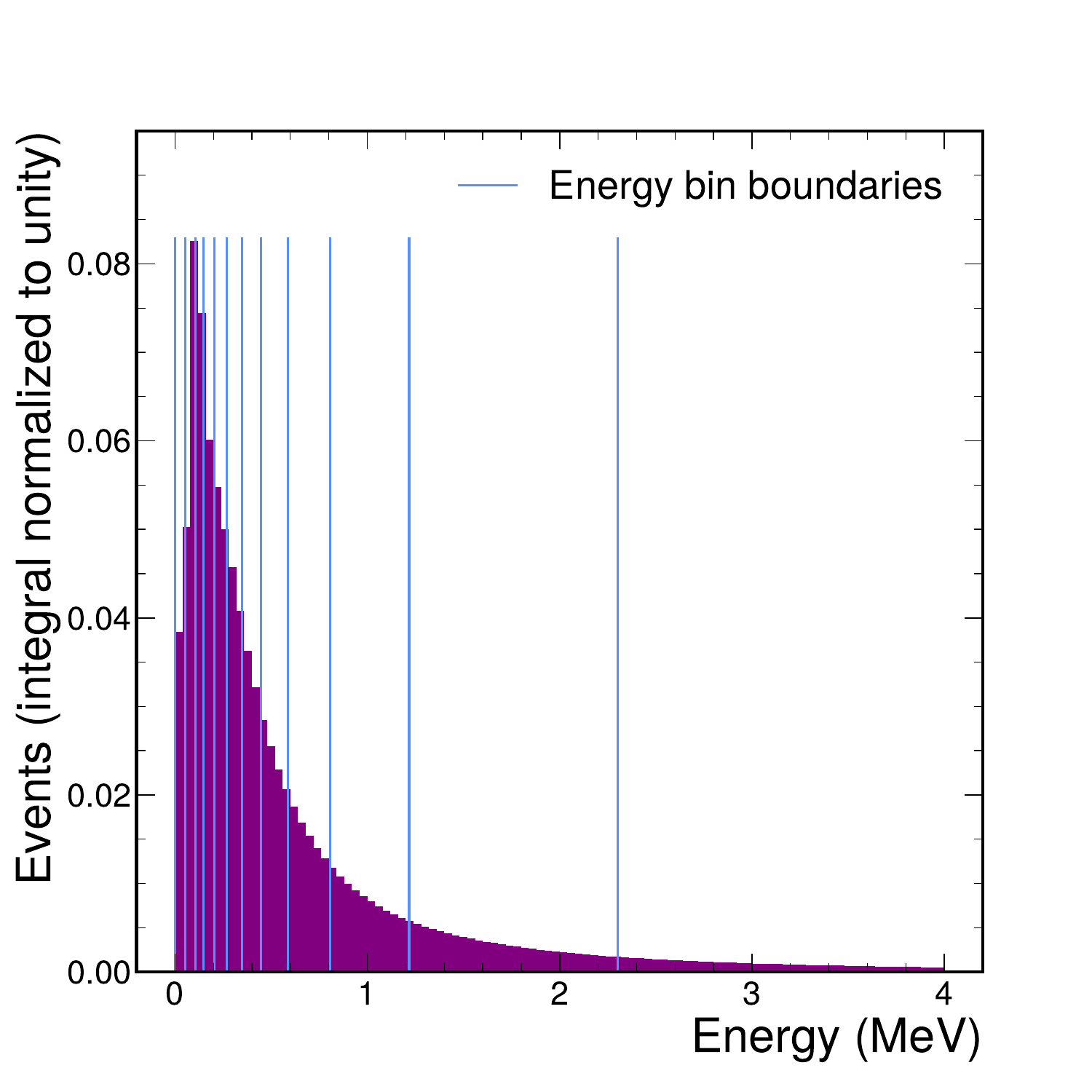}
         \caption{}
     \end{subfigure}
     \hfill
     \begin{subfigure}[b]{0.40\textwidth}
         \centering
         \includegraphics[width=\textwidth]{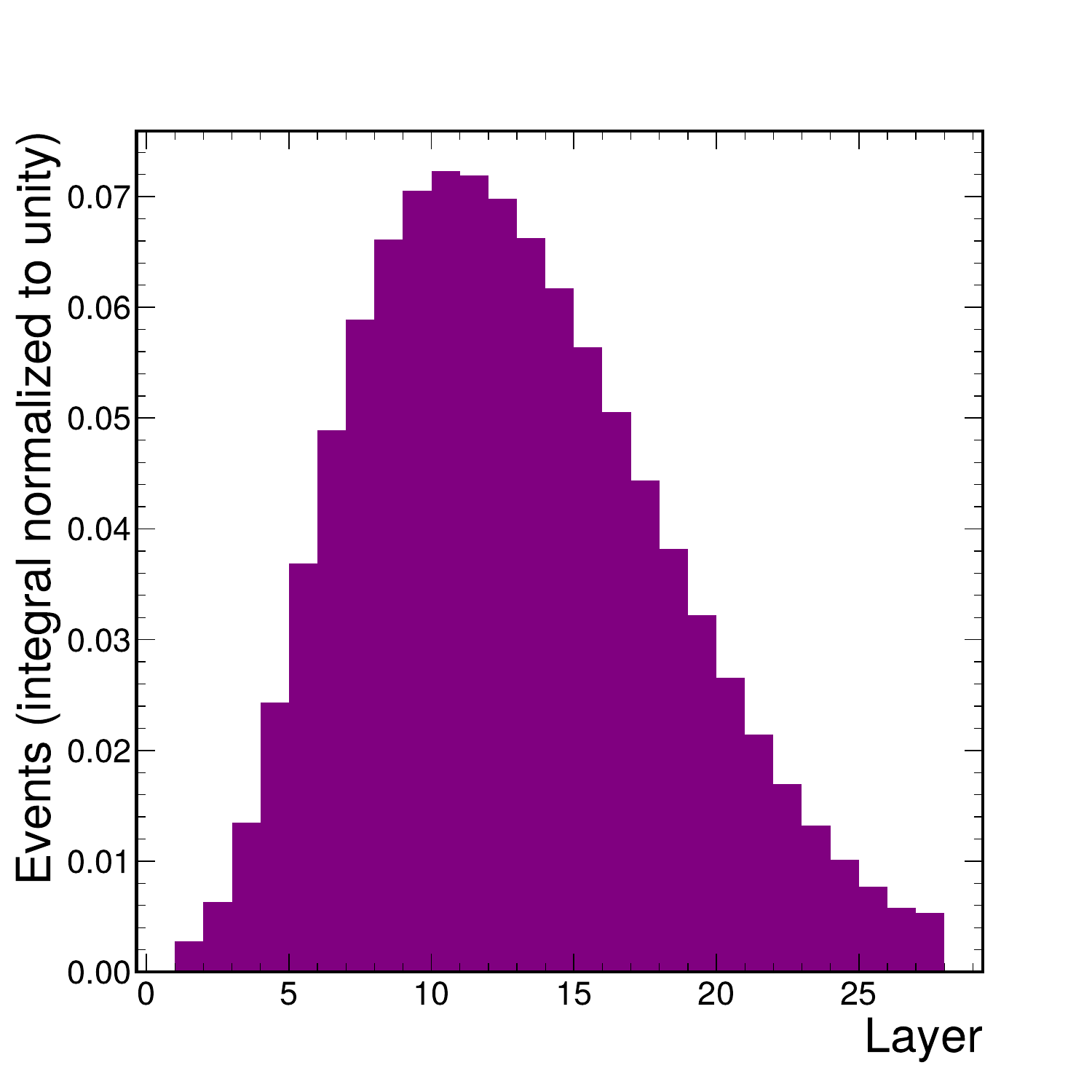}
         \caption{}
     \end{subfigure}
     \hfill\medskip
     \begin{subfigure}[b]{0.40\textwidth}
         \centering
         \includegraphics[width=\textwidth]{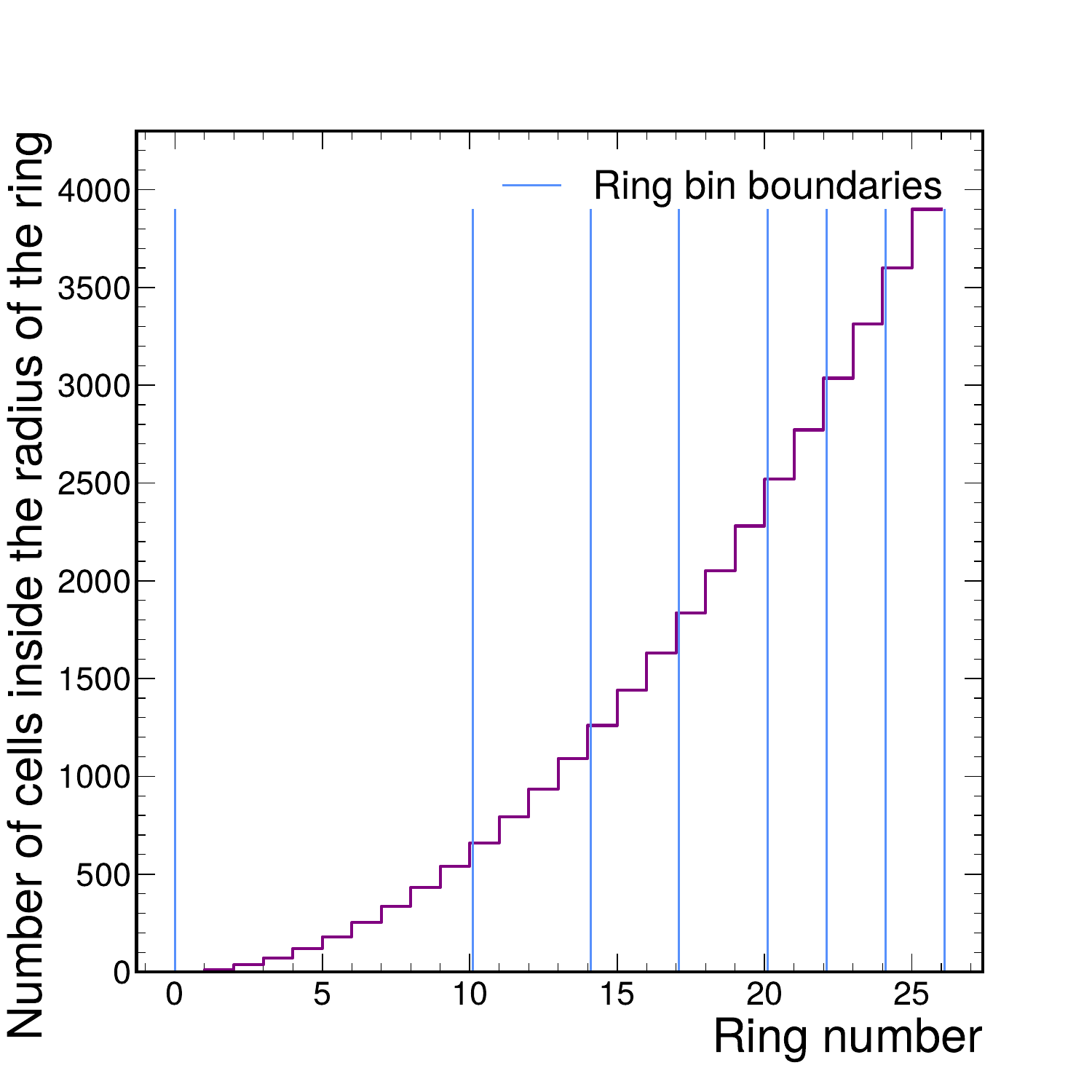}
         \caption{}
     \end{subfigure}
     \hfill
     \begin{subfigure}[b]{0.40\textwidth}
         \centering
         \includegraphics[width=\textwidth]{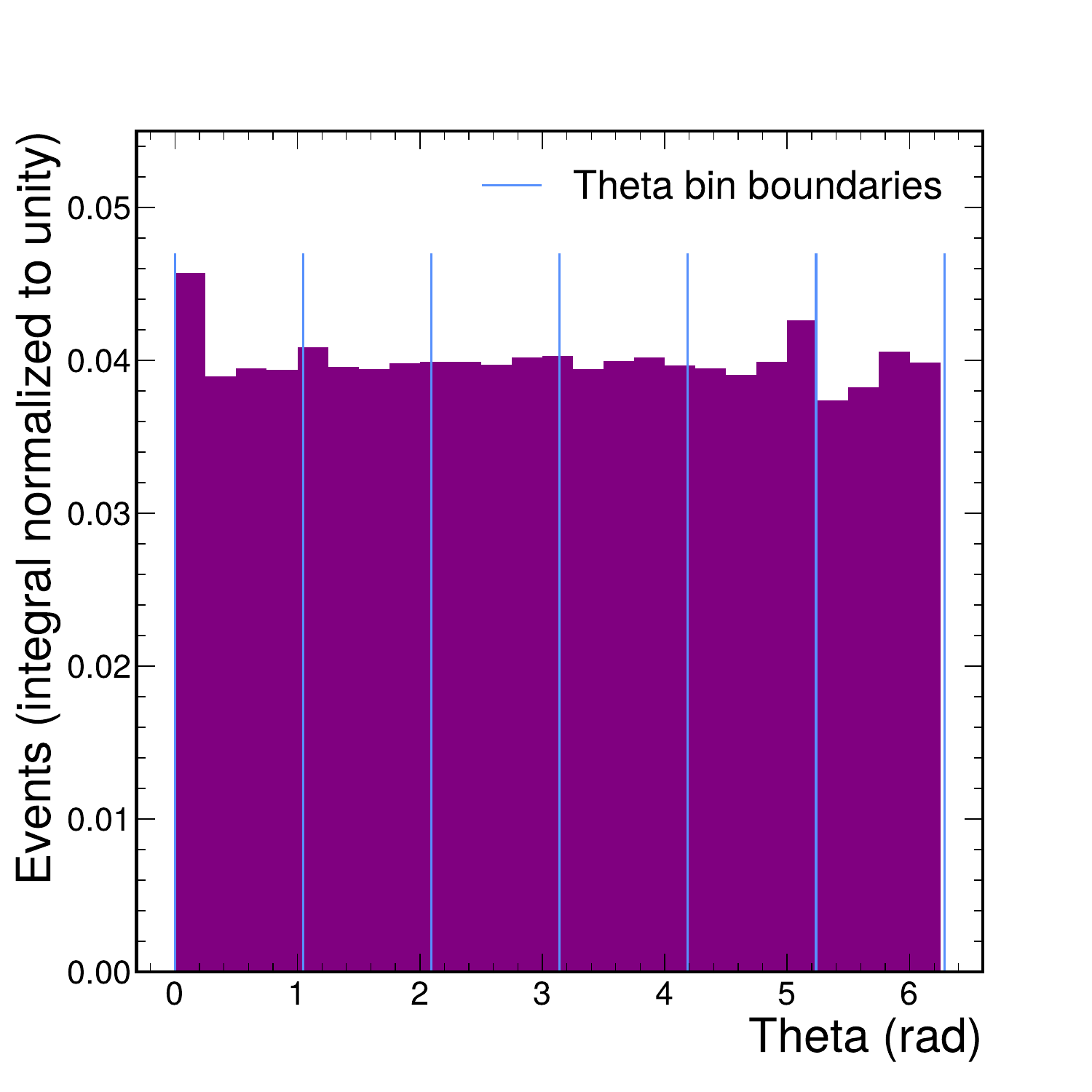}
         \caption{}
     \end{subfigure}
\caption{Distributions of the (a) cell energies asking that $S>E_\mathrm{thr}$, (b) layer number of cells satisfying $S>E_\mathrm{thr}$, (c) cumulative number of geometrical cells within the rings and (d) cell $\theta$. The bin boundaries are shown using vertical blue bars, except for the layers, where, the bins depend on the cells energies (see Eq. \ref{bins} for details).}
\label{histos}
\end{figure}

The distributions of the cells energies (corrected for the absorber), layer numbers (for the energies and layer numbers, only cells that satisfy $S_\mathrm{pos}>E_\mathrm{thr}$ are considered), geometrical ring numbers, and geometrical $\theta_k$ angles are shown in Figure \ref{histos}. In order to have sufficient statistics for the regression of the cells calibration constants, those are binned along the 3 spatial dimensions ($\ell,r,\theta_k$) and along the cell energy ($S$) direction, thus defining the following labels: $\ell_b,r_b,\theta_b,S_b$. To avoid any edge effects due to showers depositing a significant amount of their energies outside of the calorimeter volume, only electrons that are generated towards the innermost 26 rings are considered, and the calibration constants for cells between ring numbers 27 and 35 are set to unity and are not optimized. The number of bins along the $\ell,r<27,\theta_k,S$ directions is 12, 8, 6, 12 respectively. In the case of $r,\theta_k$ and $S$, the bins are chosen through a quantile method, such that the cells population in each bin is approximately equal. To account for potential deposits of energy in the hadronic section of the calorimeter, a cell in the hadronic section is considered to belong to a single geometrical layer, ring, and $\theta_k$ bin to which the labels $\ell_b=27,r_b=9,\theta_b=1$ are assigned. Its energy is still binned along the energy direction. To reflect the physical variance in the shower development in the direction of the incoming particle (longitudinal direction) as a function of the particle energy, the bins along the layer direction ($\ell$) depend on the energy bin of the cell ($S_b$). The bins edges are chosen such that, for each $S_b$ bin, the EE cells population in each $\ell_b$ bin is equal, as detailed in Eq. \ref{bins}:
\begin{equation}
\label{bins}
\ell_b\; \mathrm{bins} = 
 \begin{cases}
[1,  5,  7,  8, 10, 11, 12, 13, 15, 17, 19, 22, 26] & \text{if}~S_b=1\\
[1,  5,  7,  9, 10, 12, 13, 14, 16, 17, 19, 22, 26] & \text{if}~S_b=2\\
[1,  6,  7,  9, 10, 12, 13, 14, 16, 18, 19, 22, 26] & \text{if}~S_b=3\\
[1,  6,  7,  9, 10, 11, 13, 14, 15, 17, 19, 21, 26] & \text{if}~S_b=4\\
[1,  6,  7,  9, 10, 11, 13, 14, 15, 17, 19, 21, 26] & \text{if}~S_b=5\\
[1,  6,  7,  9, 10, 11, 12, 14, 15, 17, 19, 21, 26] & \text{if}~S_b=6\\
[1,  6,  7,  9, 10, 11, 12, 14, 15, 17, 18, 21, 26] & \text{if}~S_b=7\\
[1,  6,  7,  9, 10, 11, 12, 13, 15, 16, 18, 21, 26] & \text{if}~S_b=8\\
[1,  6,  7,  8, 10, 11, 12, 13, 14, 16, 18, 20, 26] & \text{if}~S_b=9\\
[1,  6,  7,  8,  9, 10, 11, 13, 14, 15, 17, 19, 26] & \text{if}~S_b=10\\
[1,  6,  7,  8,  9, 10, 11, 12, 13, 14, 15, 17, 26] & \text{if}~S_b=11\\
[1,  5,  6,  7,  8,  9, 10, 11, 12, 13, 14, 15, 26] & \text{if}~S_b=12    \end{cases}
\end{equation}

The procedure defines a total of 6924 distinct\footnote{$N_{\ell_b} \times N_{r_b} \times N_{\theta_b} \times N_{S_b} = 12\times 8 \times 6 \times 12=6912$ to which the 12 energy bins in the hadronic section are added.} calibration constants to be optimized by the method.

\section{Energy biases and resolution}
\label{sec:biases}

\subsection{Energy biases}
\label{subsec:biases}

The detector described in the previous section does not include any biases in the cell energy response, in the sense that every piece of material reacts the same to a given energy deposit; the geometry is very symmetrical; there is no unaccounted dead material that is typically found in physical calorimeters such as cables, cooling devices, electronics components, that can distort the energy response of one cell with respect to the others. For that reason, the detector is essentially free of biases, and calibration constants are expected to be very close to unity.\\

\begin{figure}
     \centering
     \begin{subfigure}[b]{0.30\textwidth}
         \centering
         \includegraphics[width=\textwidth]{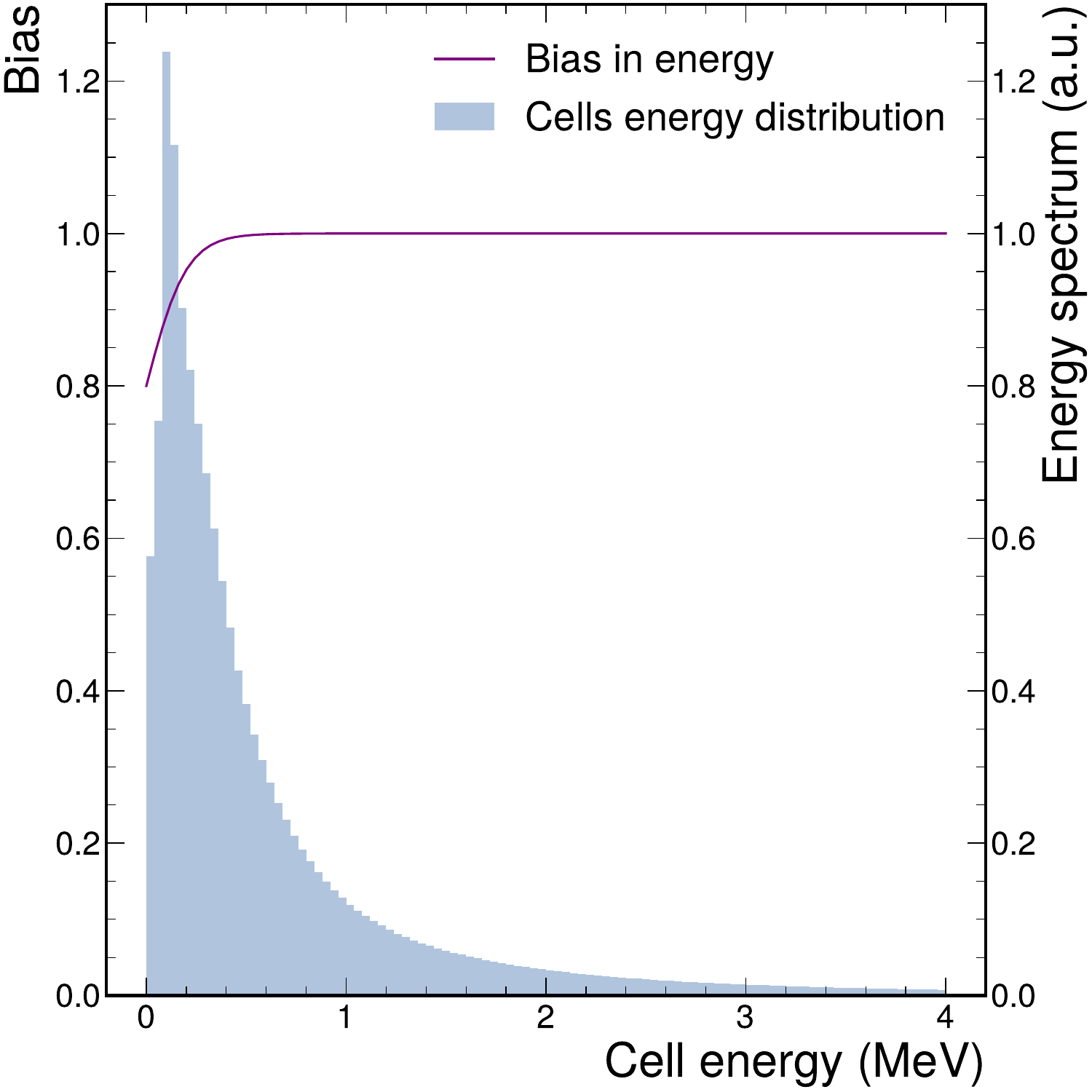}
         \caption{}
     \end{subfigure}
     \hfill
     \begin{subfigure}[b]{0.30\textwidth}
         \centering
         \includegraphics[width=\textwidth]{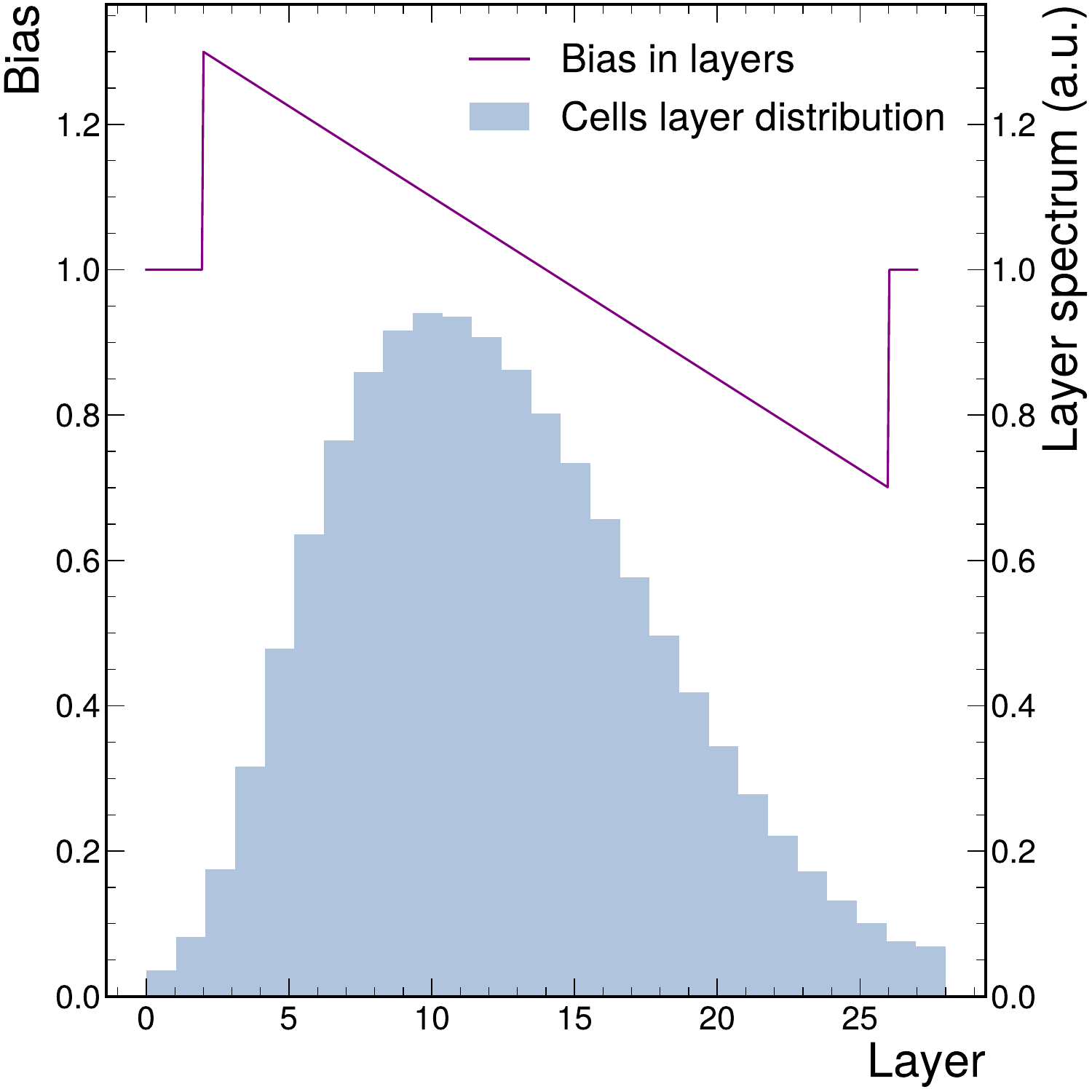}
         \caption{}
     \end{subfigure}
     \hfill
     \begin{subfigure}[b]{0.30\textwidth}
         \centering
         \includegraphics[width=\textwidth]{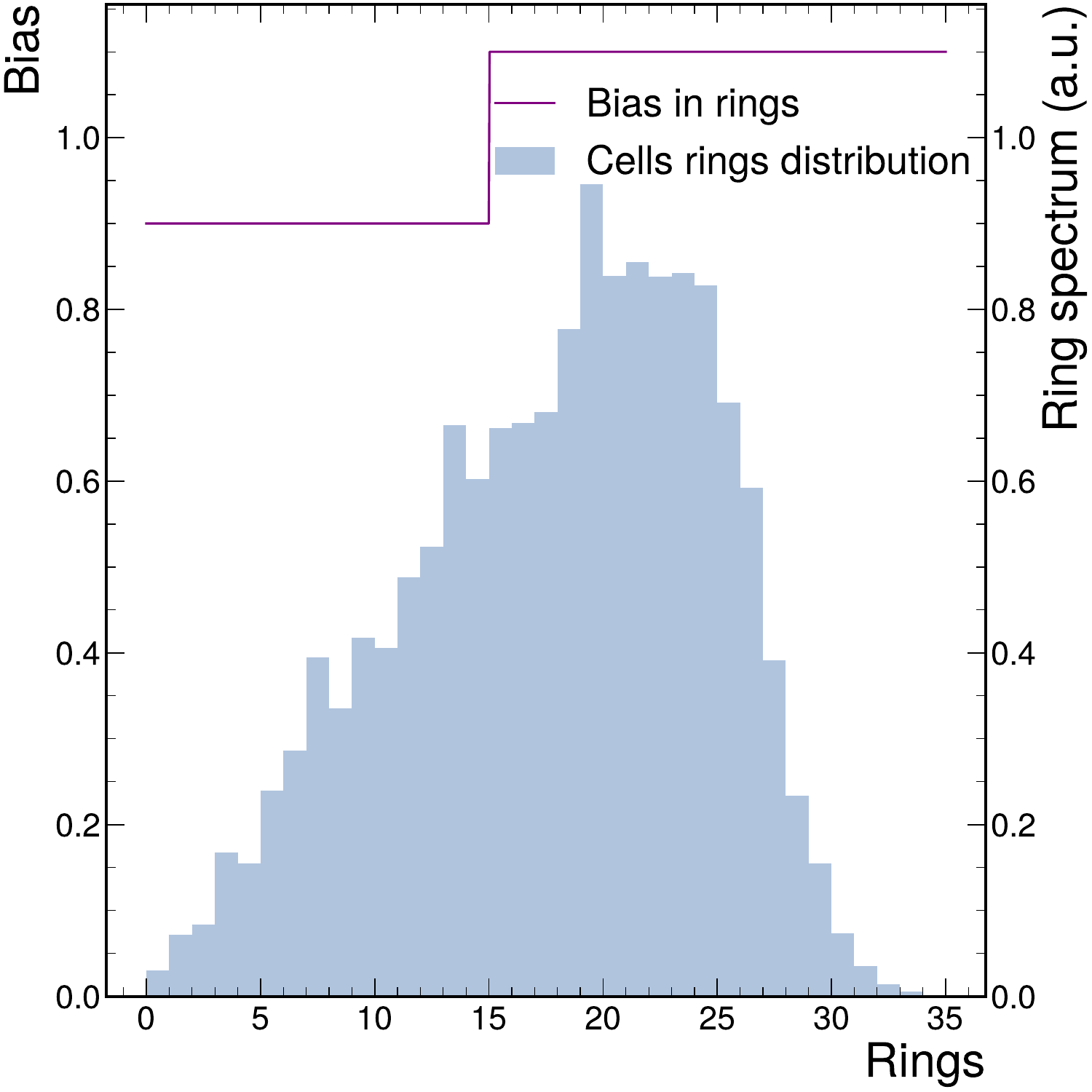}
         \caption{}
     \end{subfigure}
\caption{Colored histograms: distributions of the cells (a) energies (b) layer number and (c) ring number. Colored lines: bias functions applied to the cell energies as a function of the cell (a) energy, (b) layer and (c) ring number.}
\label{fig:biases}
\end{figure}

To emulate the response of a physical calorimeter, and hence to be able to validate the method, we introduce known biases to the cell energy response; these are taken as functions of the cell position along $\ell$ and $r$, and the cell energy ($S$). Such biases multiply the raw cell energy, and are defined in the following way:

\begin{alignat}{3}
F_S(S) =&\; 0.6 + 0.4 / (1 + e^{-10\cdot S})\\
F_\ell(\ell) = &
    \begin{cases}
       1 & \text{if}~\ell=1\\
       1.3 + 0.025 \times (\ell-2) & \text{if}~\ell\geq 2
    \end{cases}\\
F_{r}(r) = &
    \begin{cases}
       0.9 & \;\;\;\;\;\;\;\;\;\;\;\;\;\;\;\;\;\;\;\;\;\;\;\;\;\;\;\;\text{if}~r<15\\
       1.1 & \;\;\;\;\;\;\;\;\;\;\;\;\;\;\;\;\;\;\;\;\;\;\;\;\;\;\;\;\text{if}~r\geq 15
    \end{cases}
\end{alignat}
The distributions of these 3 variables, as well as the energy biases applied, are shown in Figure \ref{fig:biases}.

The Calibr-A-Ton method is then expected to compensate for such biases by optimizing the values of the calibration constants.

\subsection{Resolution effects}
\label{subsubsec:resolution_effects}

\begin{figure}[h!]
\begin{center}
\includegraphics[height=7.5cm]{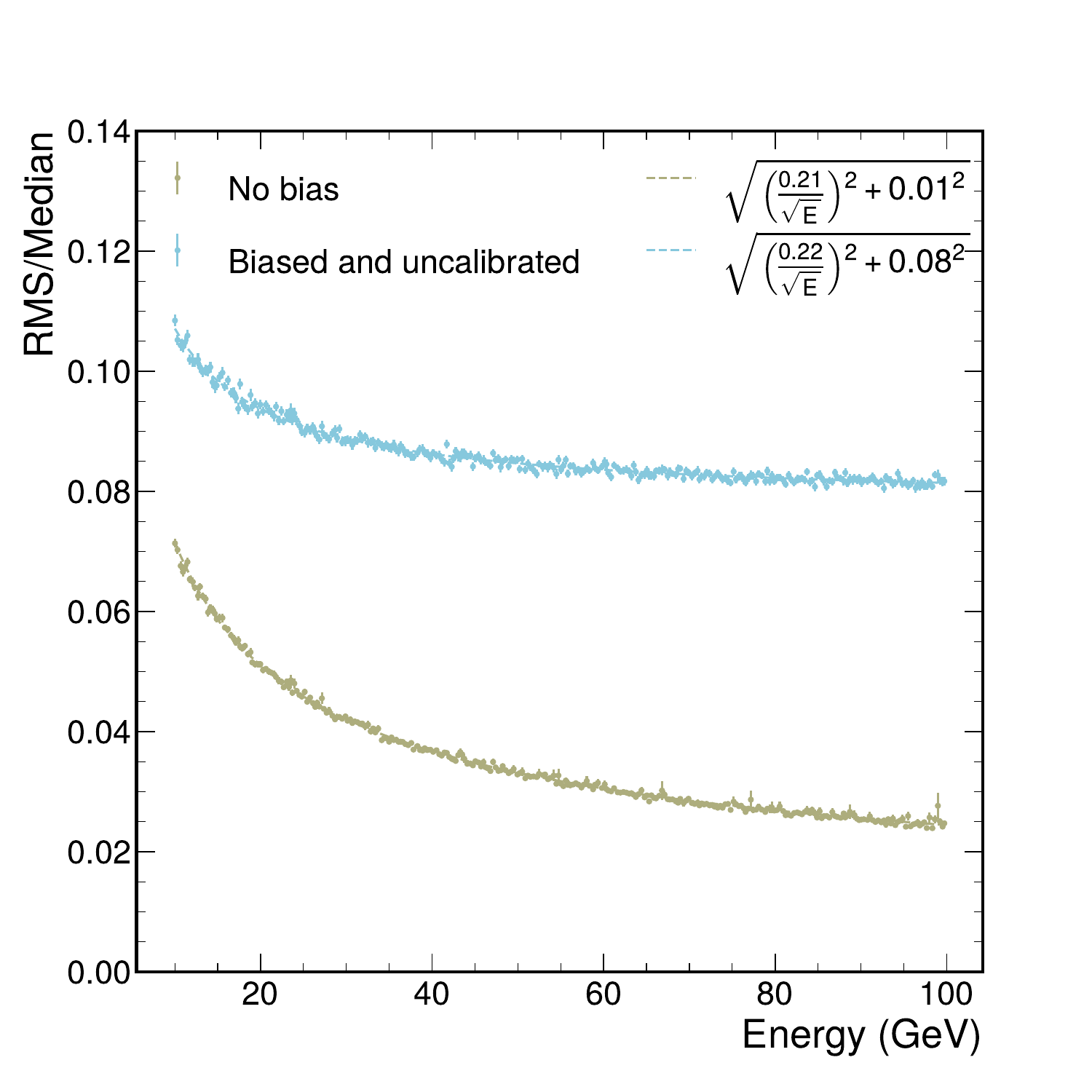}
\caption{Resolution on the shower energy, estimated by the root mean square, RMS, of the response histogram divided by the median generator energy, as a function of the generator energy. Functions of the form $f(E_\mathrm{gen})=\sqrt{a^2/E_\mathrm{gen}+c^2}$ are fitted to these points. As expected, the unbiased resolution (in dark green) presents a negligible constant term and is dominated by the stochastic term ($a=0.21\;\sqrt{\mathrm{GeV}}$). As illustrated by the blue points, the introduction of the energy biases outlined in subsection \ref{subsec:biases} leads to a substantial constant term ($c=0.08$) but leaves the stochastic term nearly unchanged ($a=0.22\;\sqrt{\mathrm{GeV}}$), as expected.}
\label{resolution_w_wo_biases}
\end{center}
\end{figure}

The energy measured of a calorimeter is affected by resolution effects. These can come from three sources: statistical fluctuations in the number of charges produced by the shower and/or collected by the cell (also known as the \emph{stochastic term}); noise due to the electronics or other physical processes such as pileup collisions (also known as the \emph{noise term}); and imperfections in the shower containment, material inhomogeneities, variations in the dimensions in the calorimeter, unaccounted material in front of or within the calorimeter, etc. (collectively known as the \emph{constant term}). The relative energy resolution $\sigma_E/E$ is usually parametrized as:
\begin{equation}    
\frac{\sigma_E}{E} = \frac{a}{\sqrt{E}} \oplus \frac{b}{E} \oplus c,
\label{eq:resolution}
\end{equation}
where the terms associated with $a,b,c$ represent the stochastic, noise and constant terms, respectively, that are added in quadrature. Only the stochastic component is modeled through \texttt{GEANT4} in our simulation. The effect of the noise term on the electromagnetic shower resolution in HGCAL has been estimated to be negligible \cite{HGCALtestbeam} and for that reason is not included here either ($b=0$). The shower energy resolution in the presence or the absence of energy biases is shown in Figure \ref{resolution_w_wo_biases}. The points are fitted with functions of the form of the one of Eq. \ref{eq:resolution}. Thanks to the large statistics involved in the fits, the relative uncertainties on the $a,c$ parameters extracted are always less one the per-mil and are therefore not reported. The unbiased case shows that the simulation is already affected by a constant term ($c=0.01$), albeit negligible compared to the stochastic term ($a=0.21\;\sqrt{\mathrm{GeV}}$) in the energy range considered. The fact that the constant term is much amplified ($c=0.08$) when an energy bias, of the form described in subsection \ref{subsec:biases}, is applied, means that the energy bias acts as an additional source of uncertainty on the shower energy measurement. In this paper, we assess the capability of the method to mitigate the impact of biases, such that the energy resolution after calibration, in the presence of biases, is as close as possible to the uncalibrated resolution, in the absence of biases.

\section{Performance}
\label{performance}

\begin{figure}[h!]
\begin{center}
\includegraphics[height=6cm]{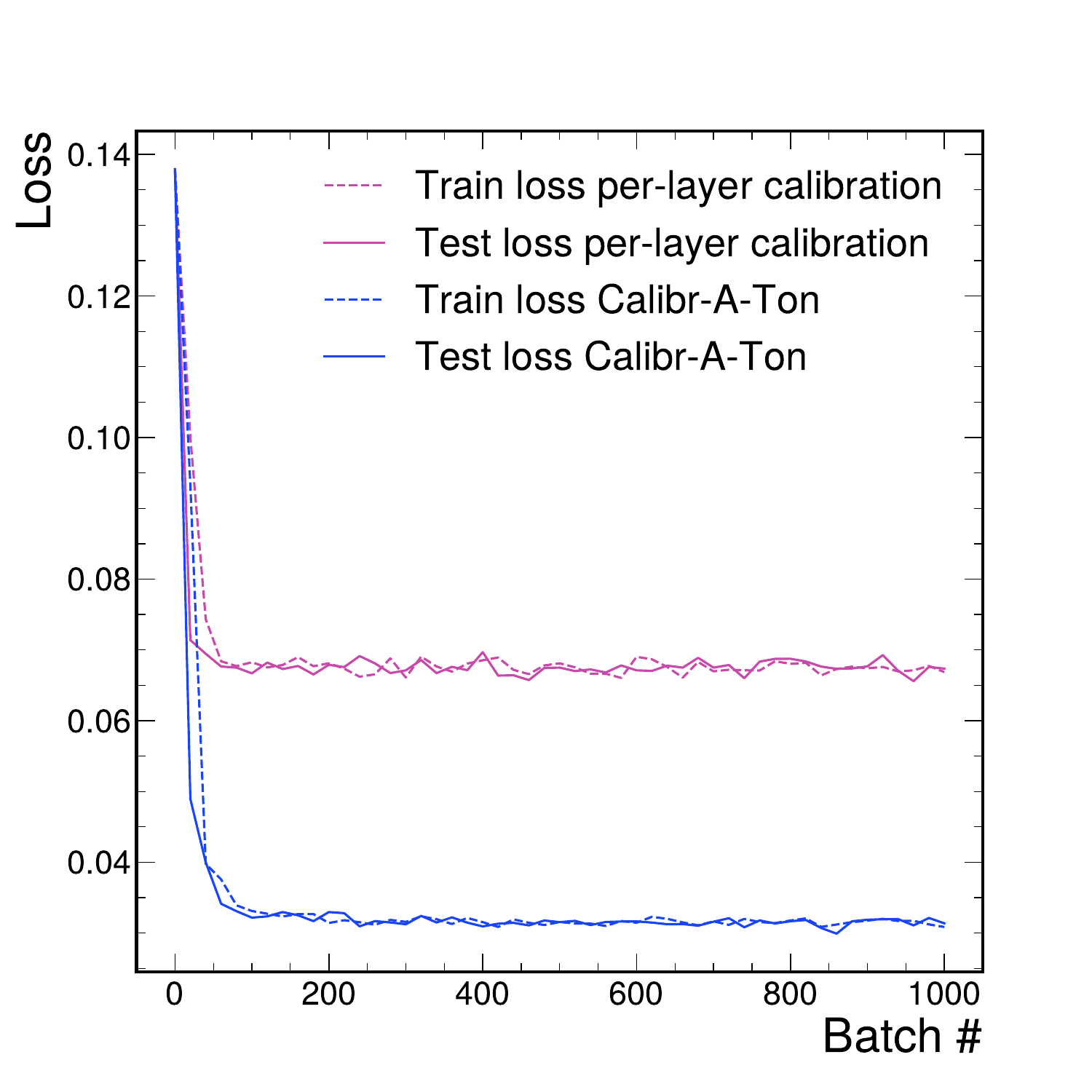}
\includegraphics[height=6cm]{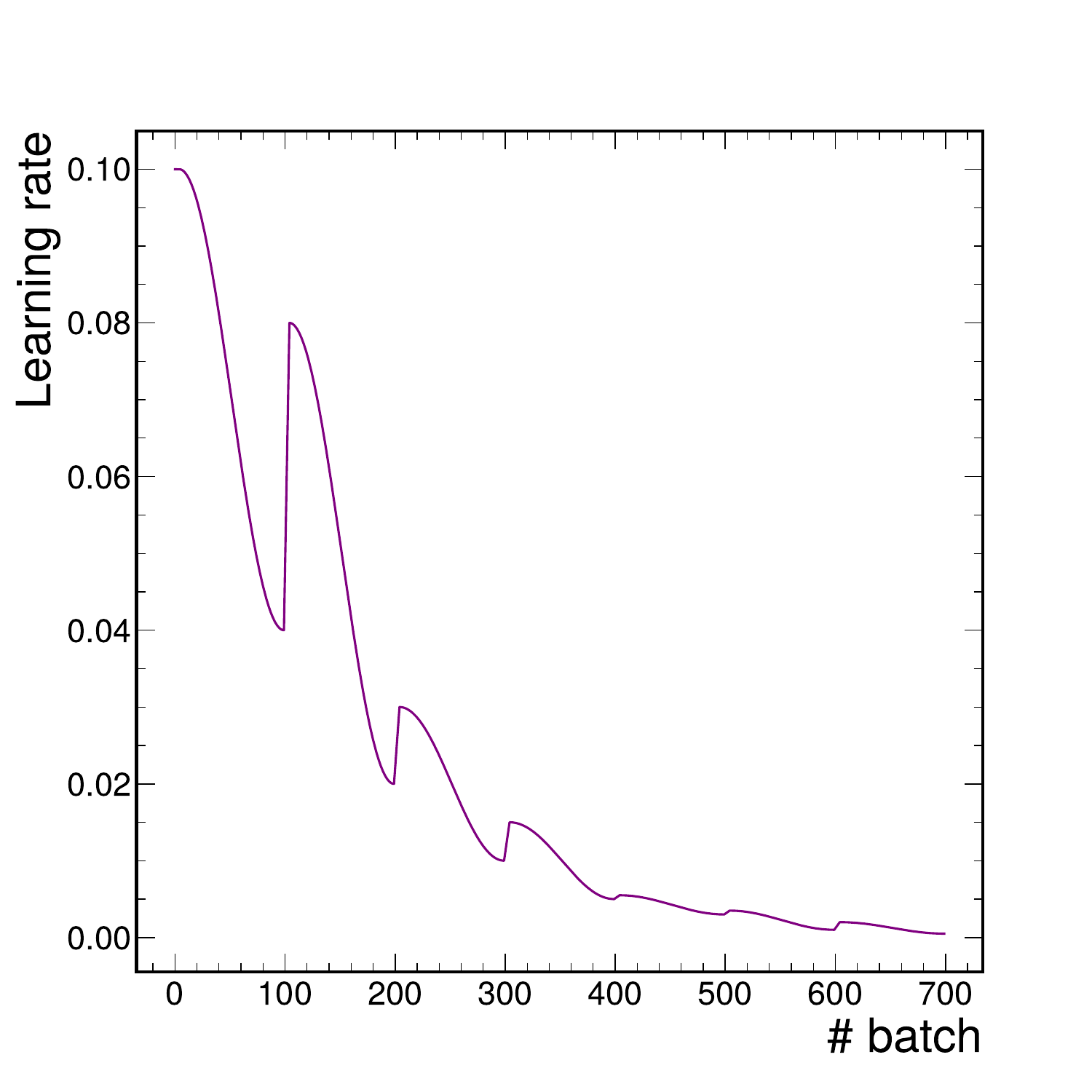}
\caption{(Left) Value of the loss function of Eq. \ref{eq:loss} as a function of the batch number for the per-layer calibration (in magenta) and for Calibr-A-Ton (in blue). The training losses (dashed line) are evaluated as the average of the losses of the previous 14 training batches or 14 weight updates, corresponding in total to 7000 training electrons. The testing losses (solid line) are computed once every 14 batches. (Right) Value of the learning rate vs. the batch number, a process called ``annealing''. The same learning rate values are used for either methods.}
\label{fig:loss}
\end{center}
\end{figure}

The method described in Section \ref{sec:desc}, and the geometry and data sample described in Section \ref{sec:calo}, are used to extract the calibration constants described in subsection \ref{subsec:binning} for either methods (per-layer calibration and Calibr-A-Ton) in the presence of the biases introduced in subsection \ref{subsec:biases}. In this section, the resulting performance coming from either method are compared. The input dataset is split between a training sample, containing $N_\mathrm{train}=3.5\times 10^5$ electrons, and a testing sample, containing $N_\mathrm{test}=1.5 \times 10^5$ electrons. The samples are further split into batches of size $N=500$ electrons. The calibration constants are all initialized at 1.0. The \texttt{JAX} open-source library \cite{JAX} is used to perform an automatic differentiation of a function encoding the loss function of Eq. \ref{eq:loss}. The value of the loss function as a function of the batch number is shown for either method on Figure \ref{fig:loss} (left).

\begin{figure}[h!]
\begin{center}
\includegraphics[height=7.5cm]{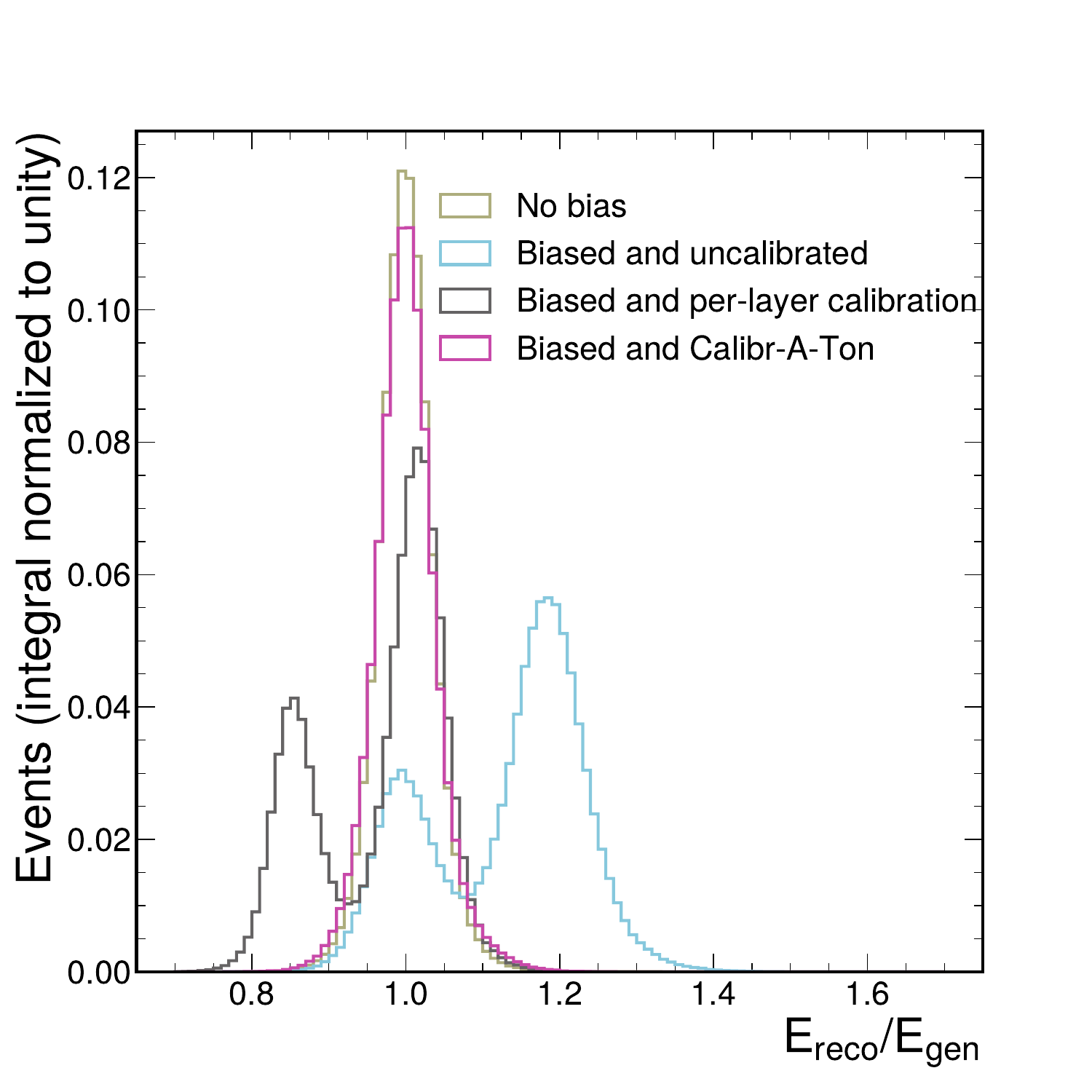}
\caption{Histograms of the response, as defined in Eq. \ref{eq:response}, for electron showers, when: no bias is introduced (green); the bias is applied but no calibration constants are derived (light blue); the bias is applied and the calibration constants from the per-layer calibration are included (gray); the bias is applied and the calibration constants from Calibr-A-Ton are included (magenta).}
\label{fig:inclusive_response}
\end{center}
\end{figure}

\begin{figure}[htp]
\centering
\includegraphics[width=.3\textwidth]{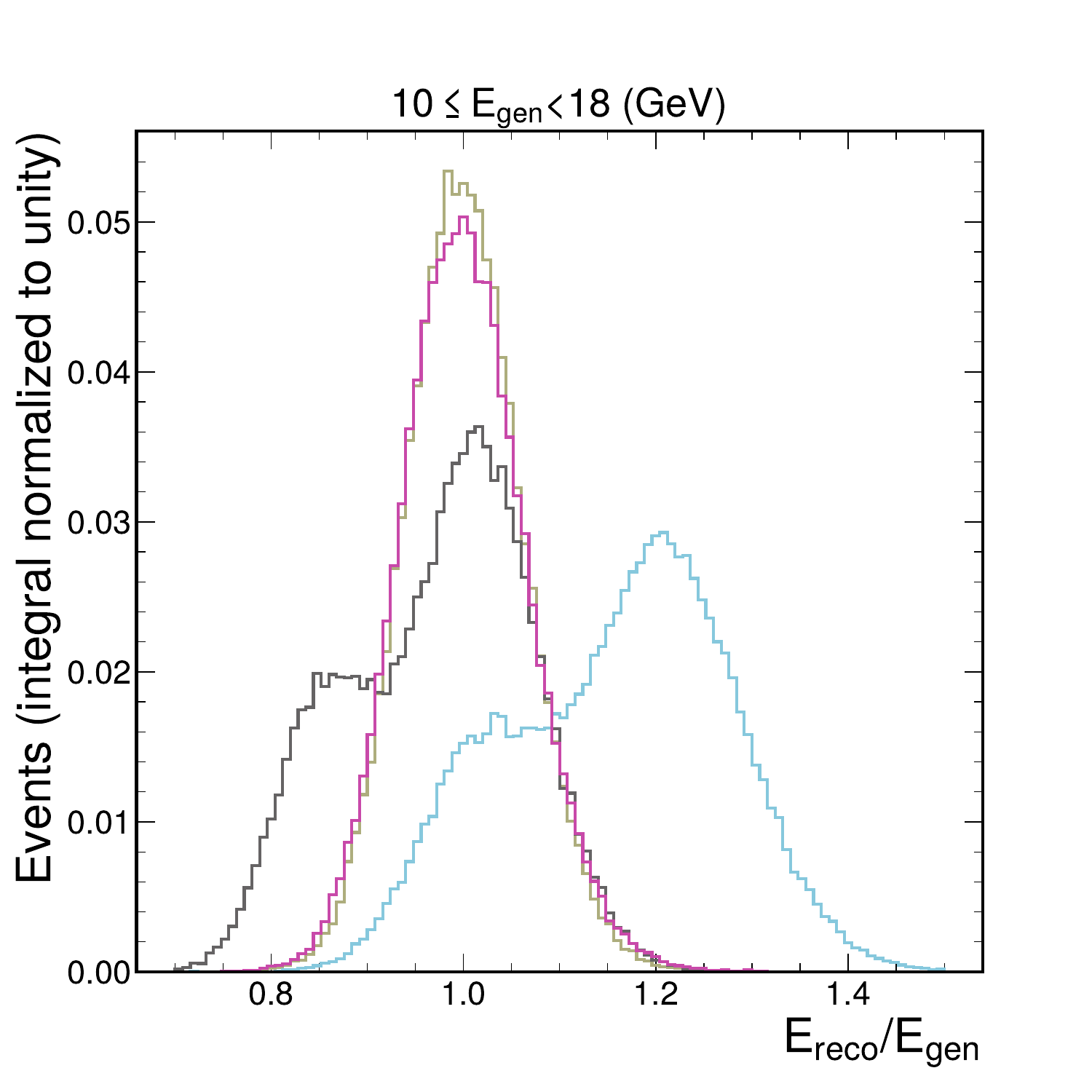}\quad
\includegraphics[width=.3\textwidth]{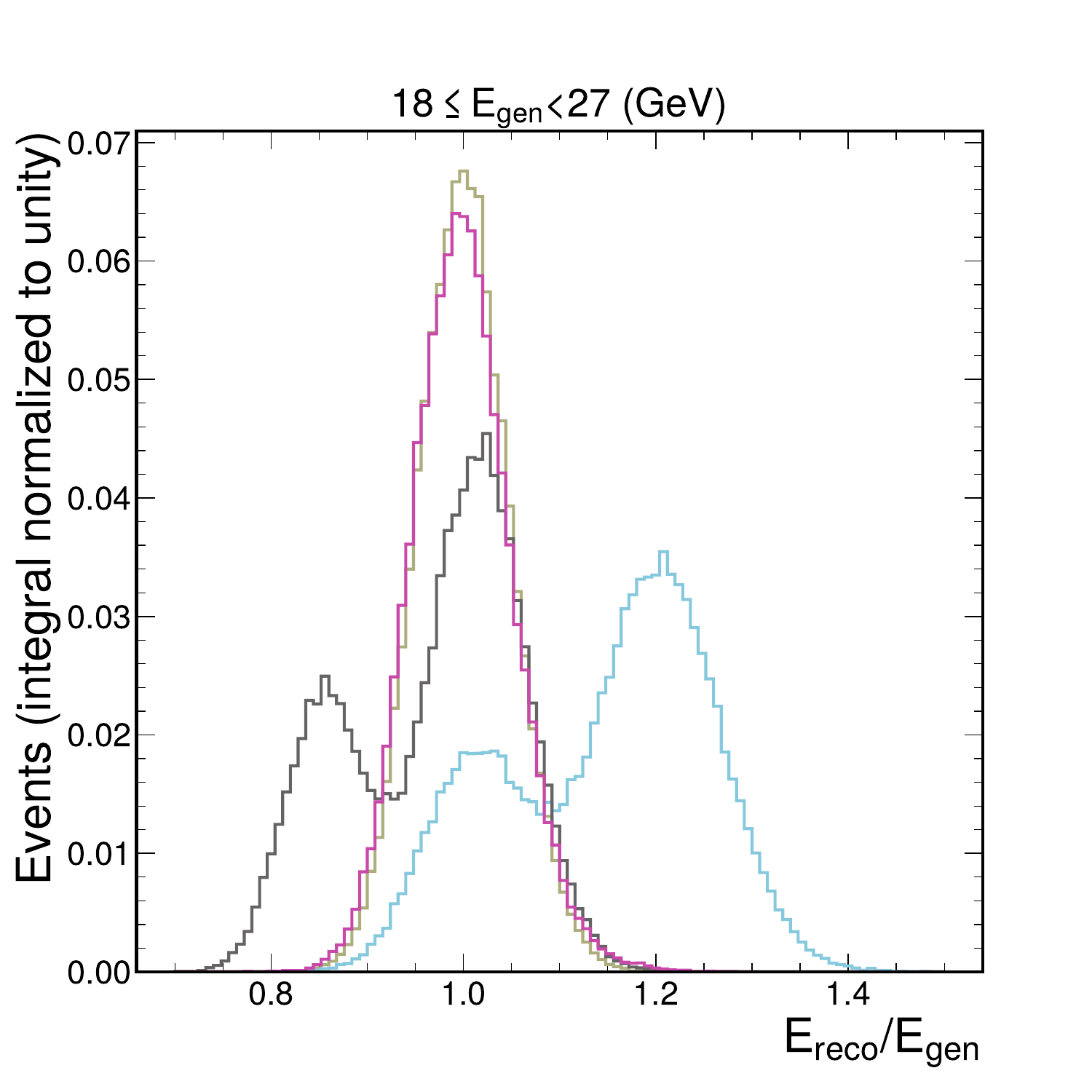}\quad
\includegraphics[width=.3\textwidth]{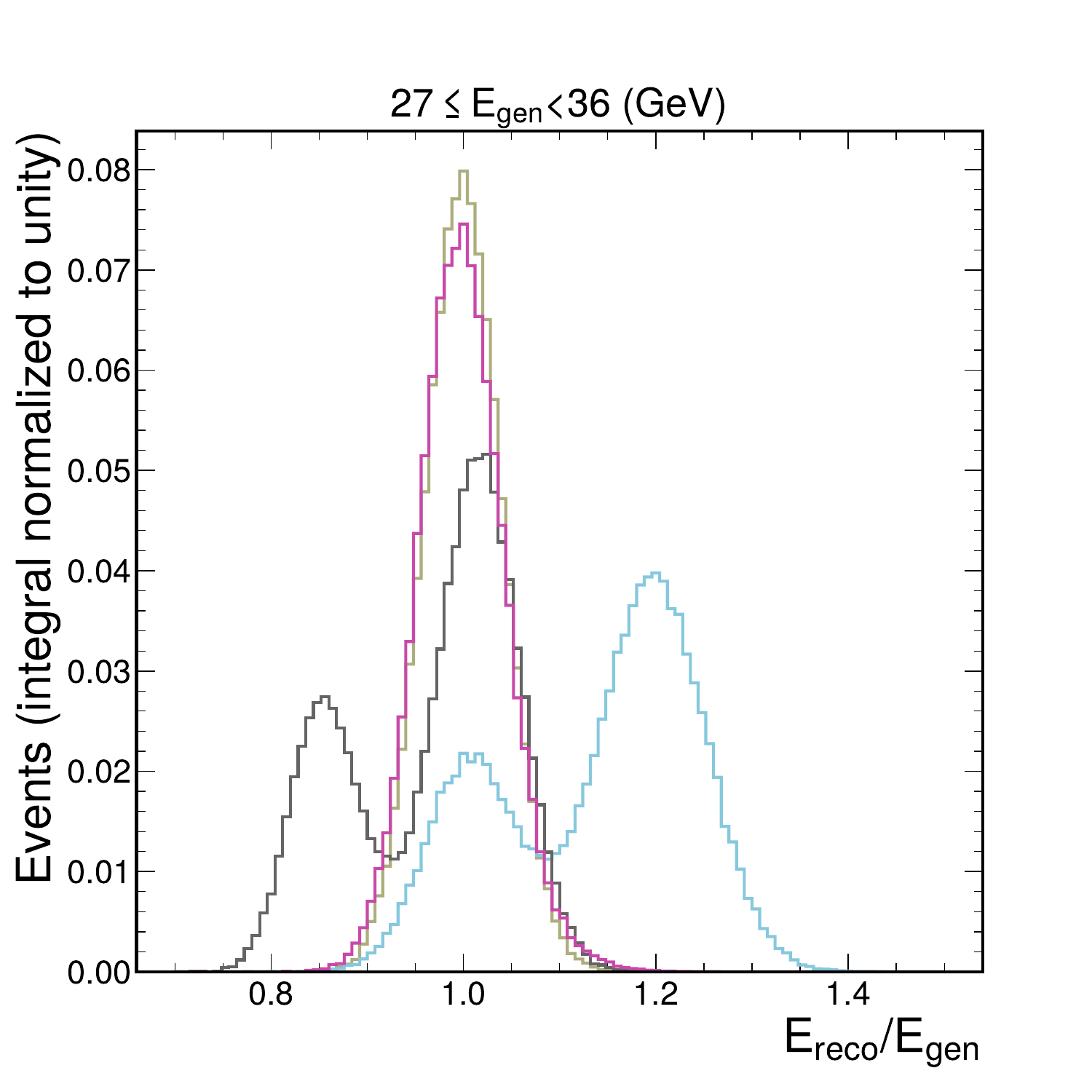}

\medskip

\includegraphics[width=.3\textwidth]{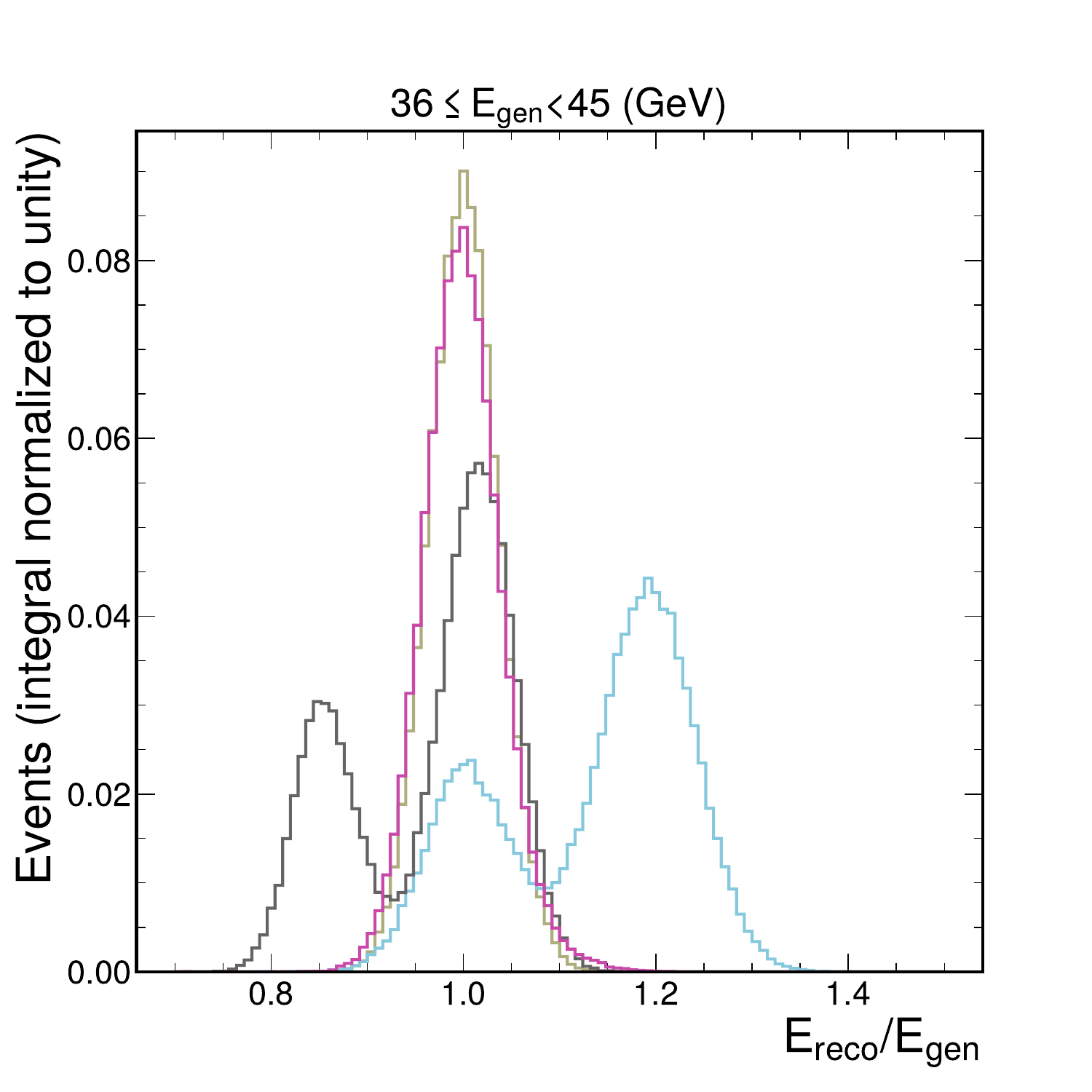}\quad
\includegraphics[width=.3\textwidth]{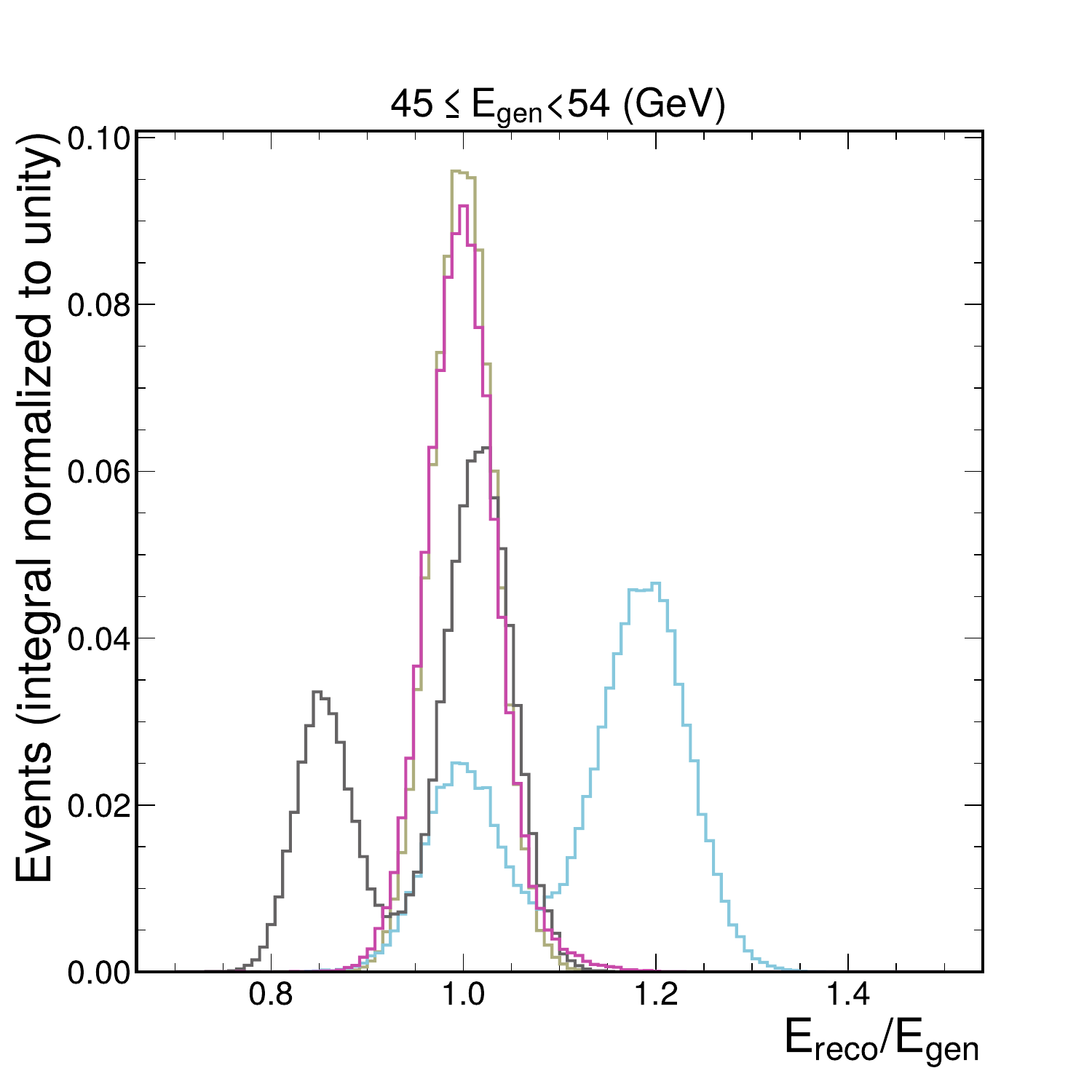}\quad
\includegraphics[width=.3\textwidth]{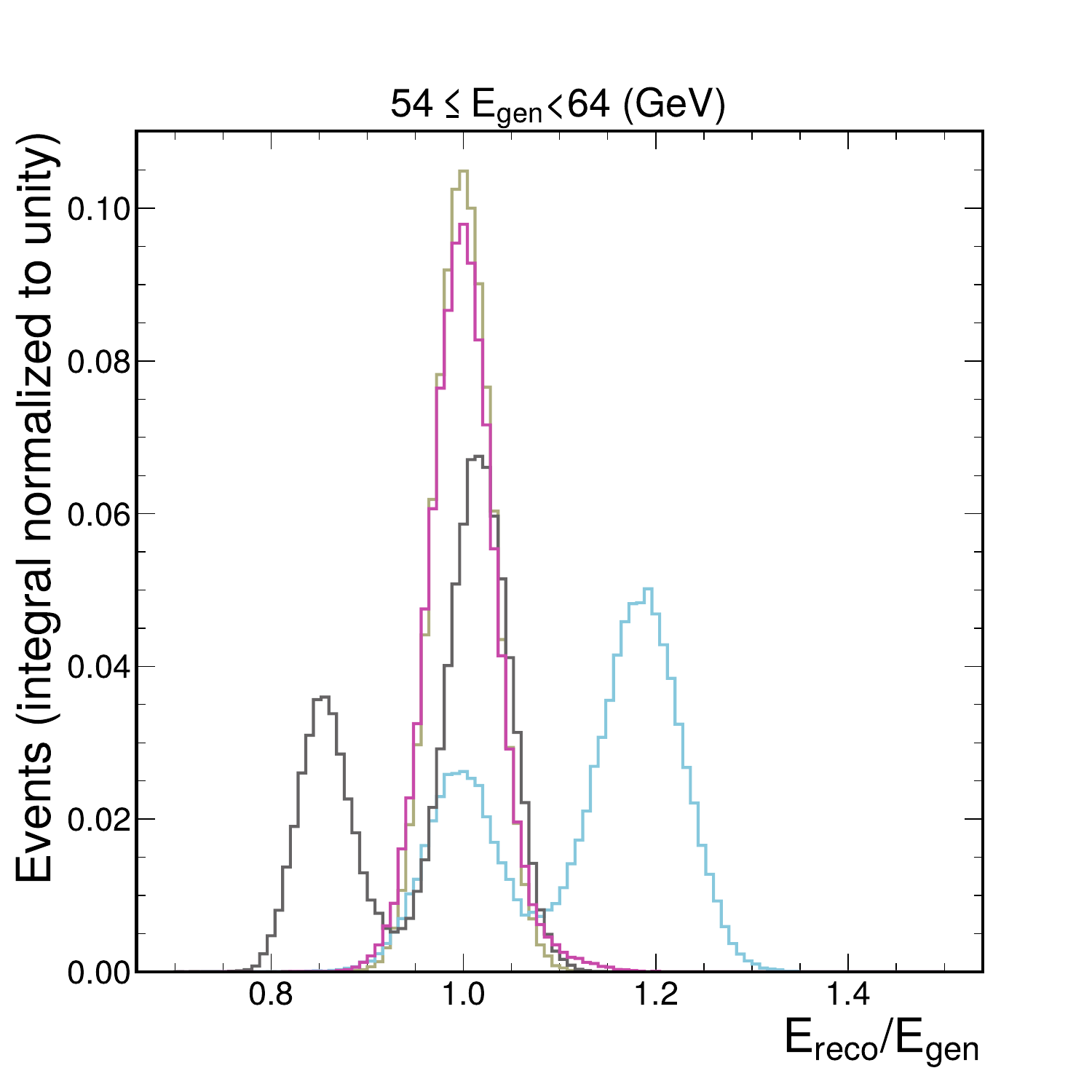}

\medskip

\includegraphics[width=.3\textwidth]{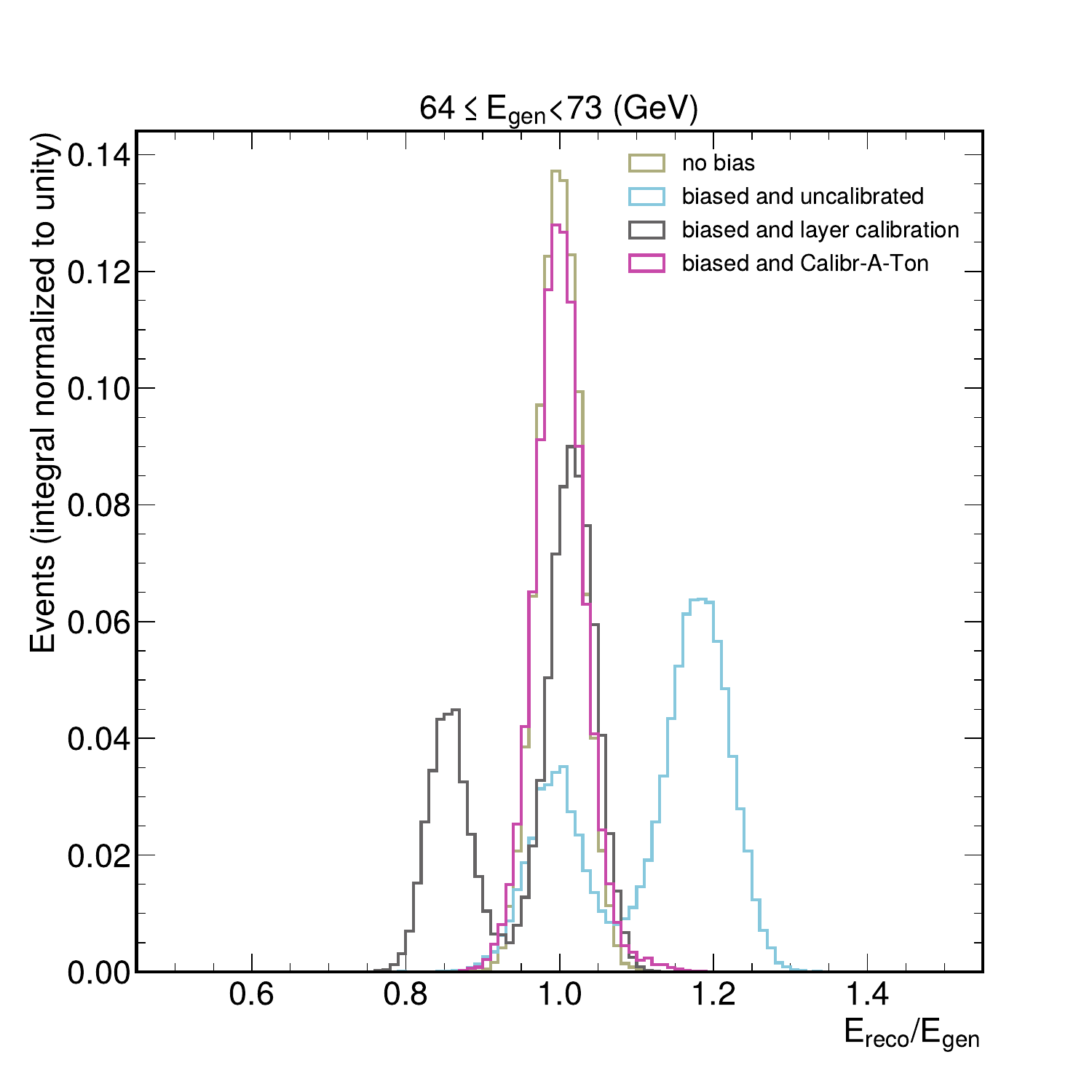}\quad
\includegraphics[width=.3\textwidth]{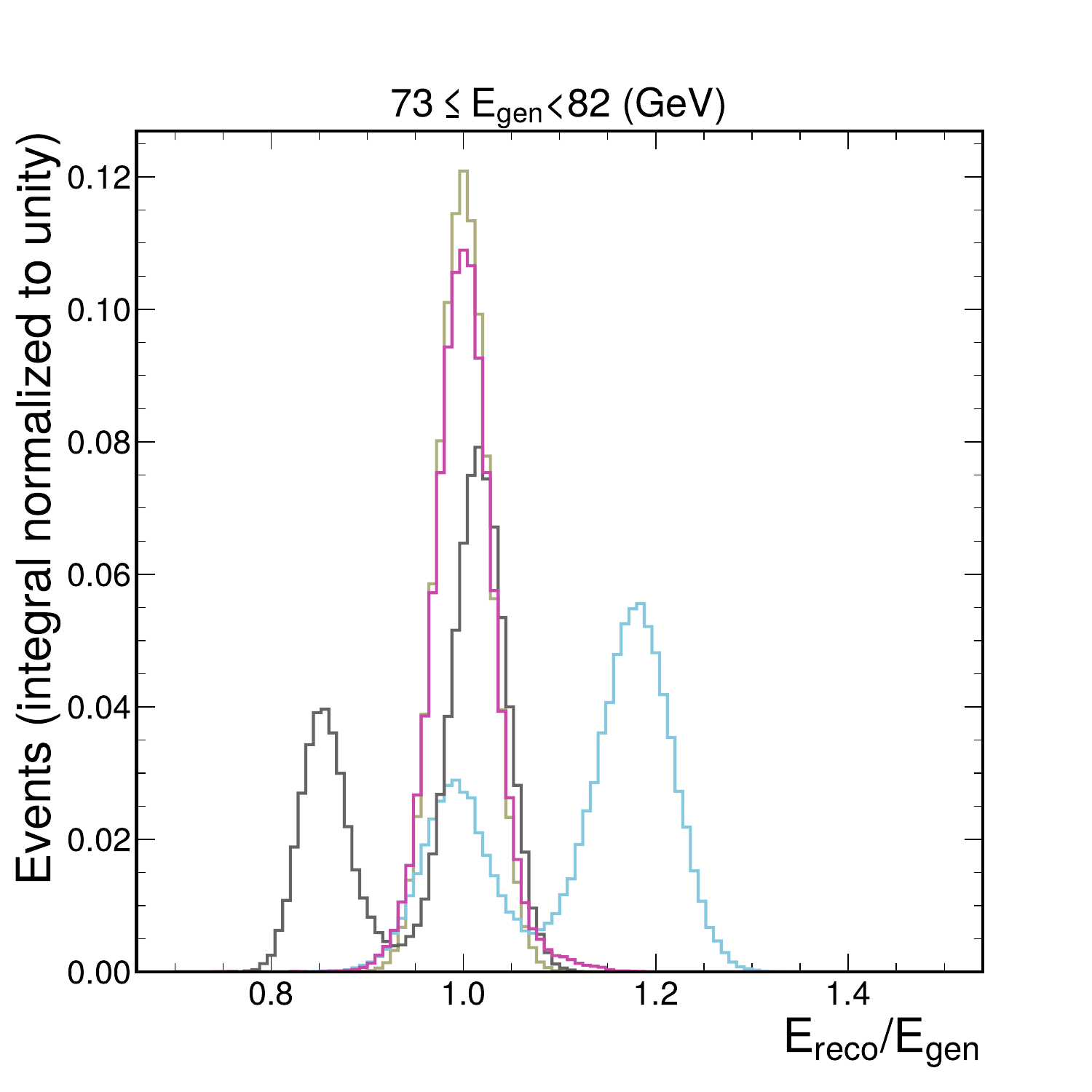}\quad
\includegraphics[width=.3\textwidth]{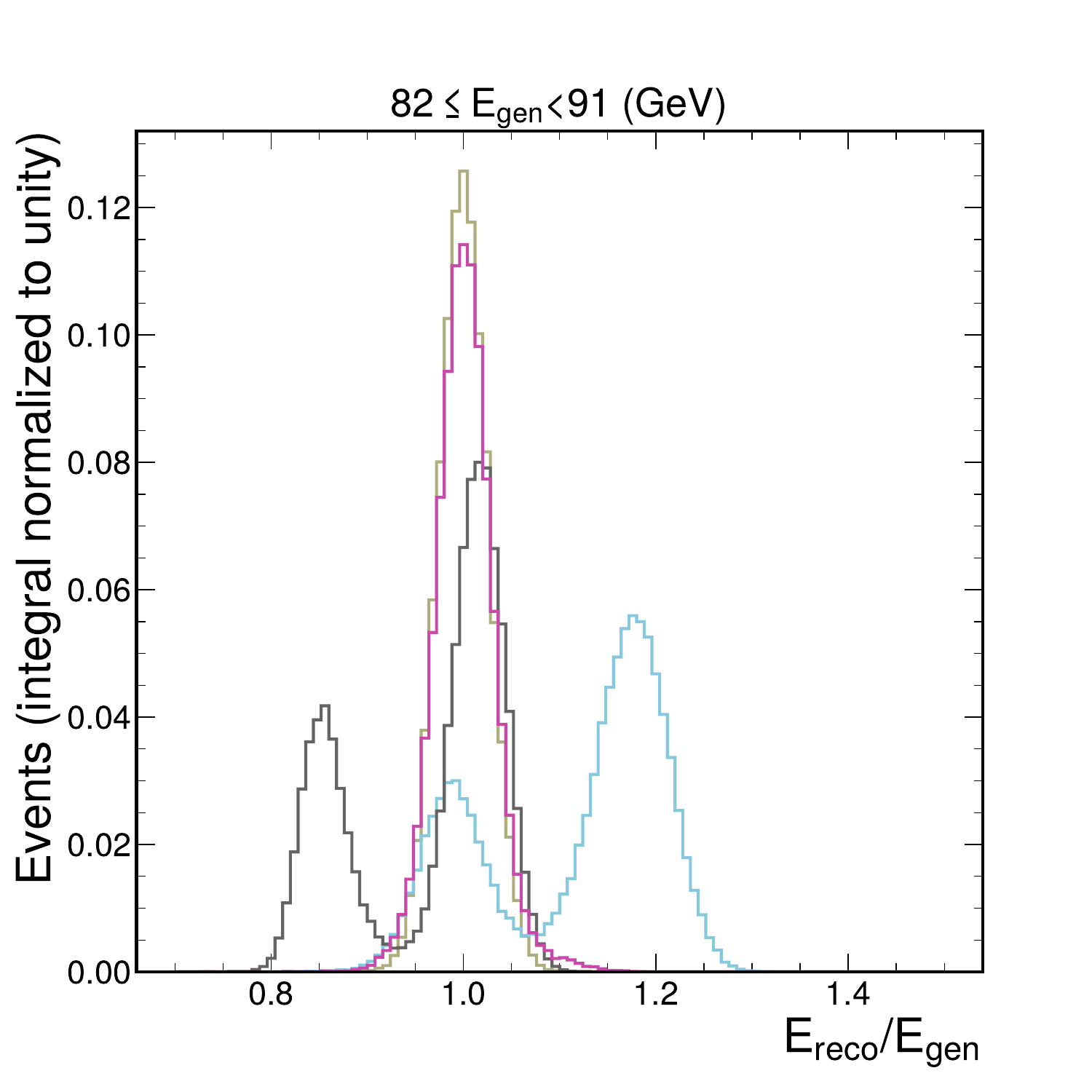}

\medskip

\includegraphics[width=.3\textwidth]{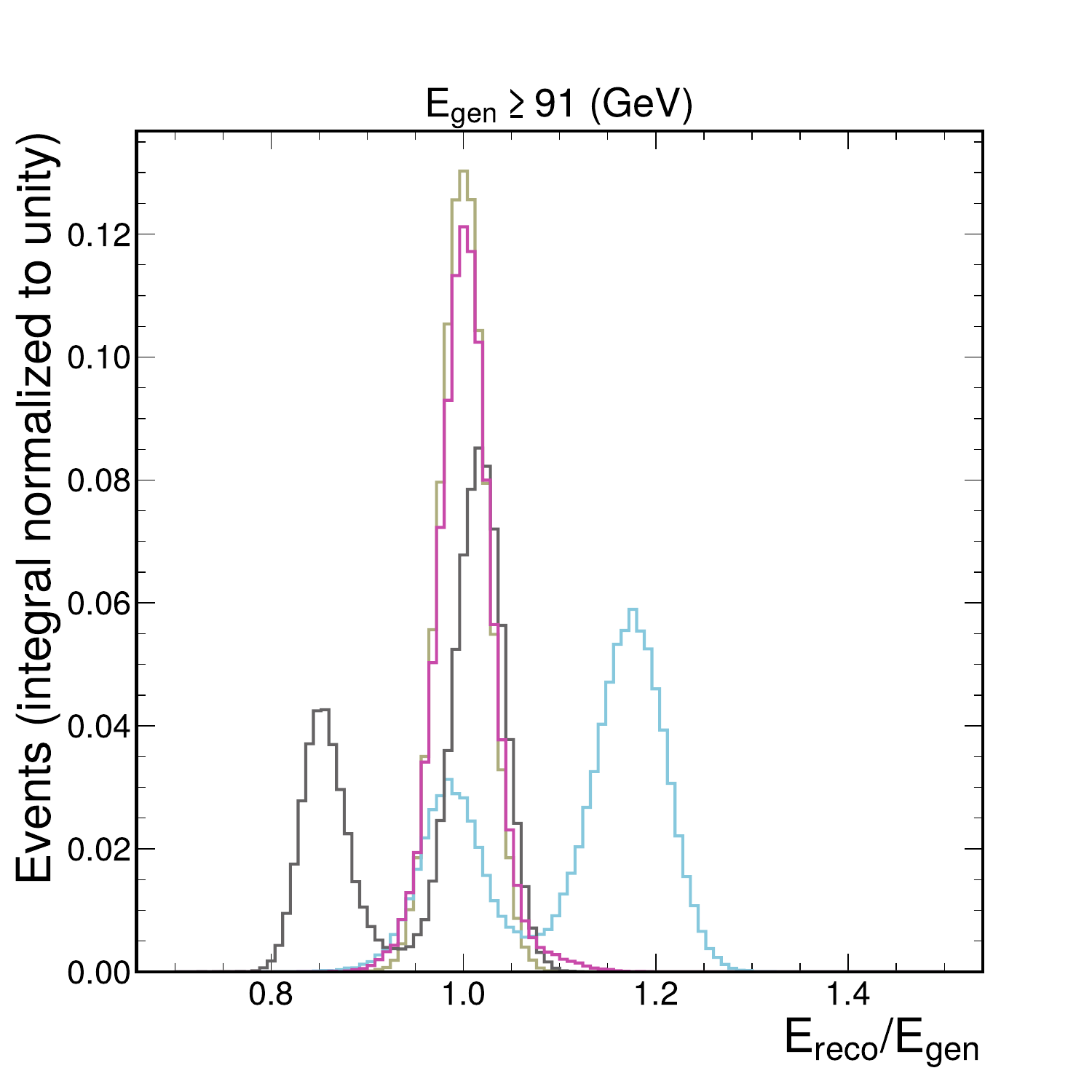}
\includegraphics[width=.3\textwidth]{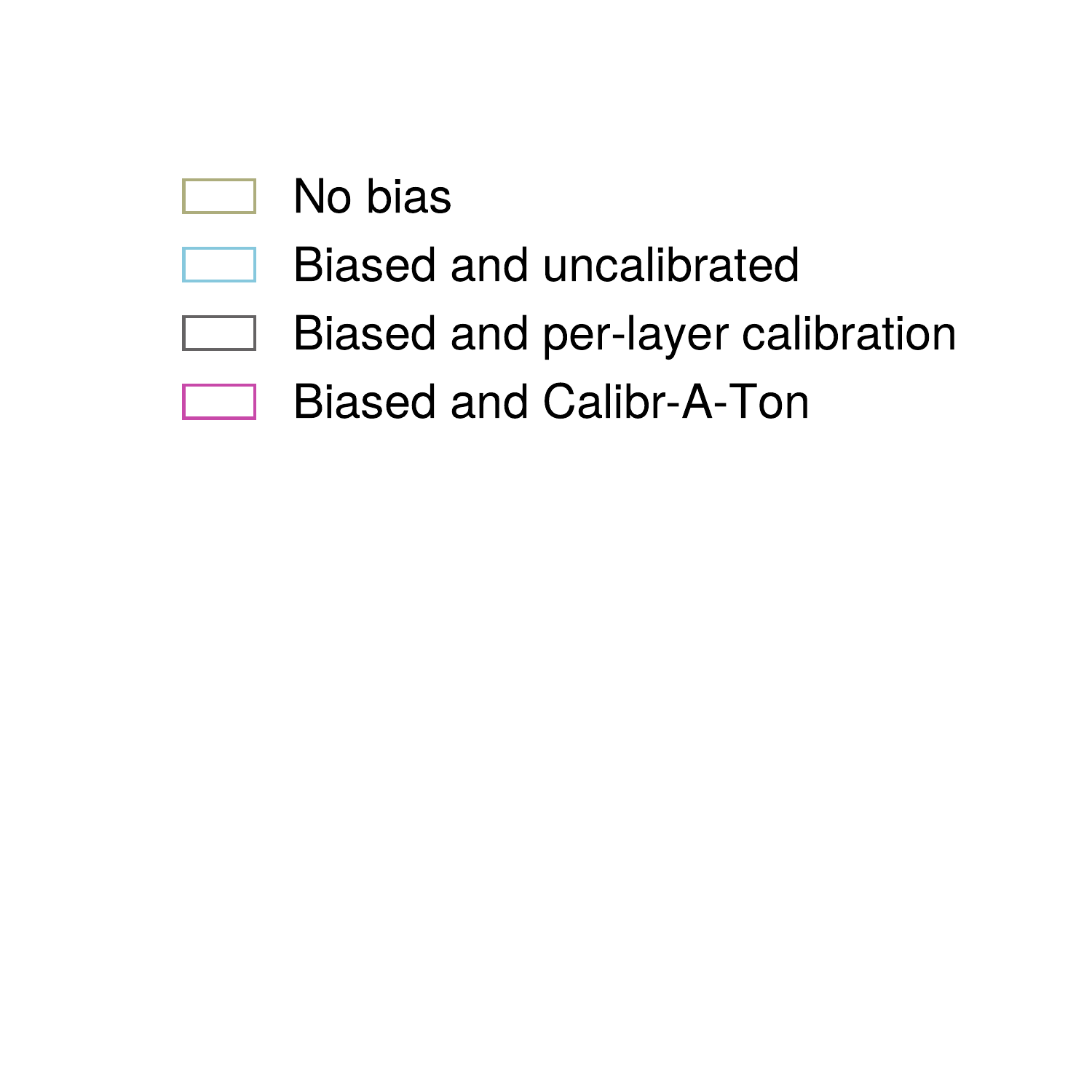}

\caption{Distributions of the response, as defined in Eq. \ref{eq:response}, for electron showers (the absorber correction is included), in bins of the generator electron energy.}
\label{fig:responses_binned}
\end{figure}

The distributions of the response, as defined in Eq. \ref{eq:response}, for the inclusive testing sample are shown for the different conditions and for the different training methods in Figure \ref{fig:inclusive_response}. With respect to the raw response of the calorimeter, the application of the biases distorts the response, and in particular generates a double-peak structure, which is not absorbed using the per-layer calibration. On the other hand, the calibration constants derived with the Calibr-A-Ton method allow for an almost full absorption of the bias, providing a response very similar to the situation without bias. This is confirmed when looking at the response in bins of the generator electron energy, as shown in Figure \ref{fig:responses_binned}.\\

\begin{figure}[h!]
\begin{center}
\includegraphics[height=7.5cm]{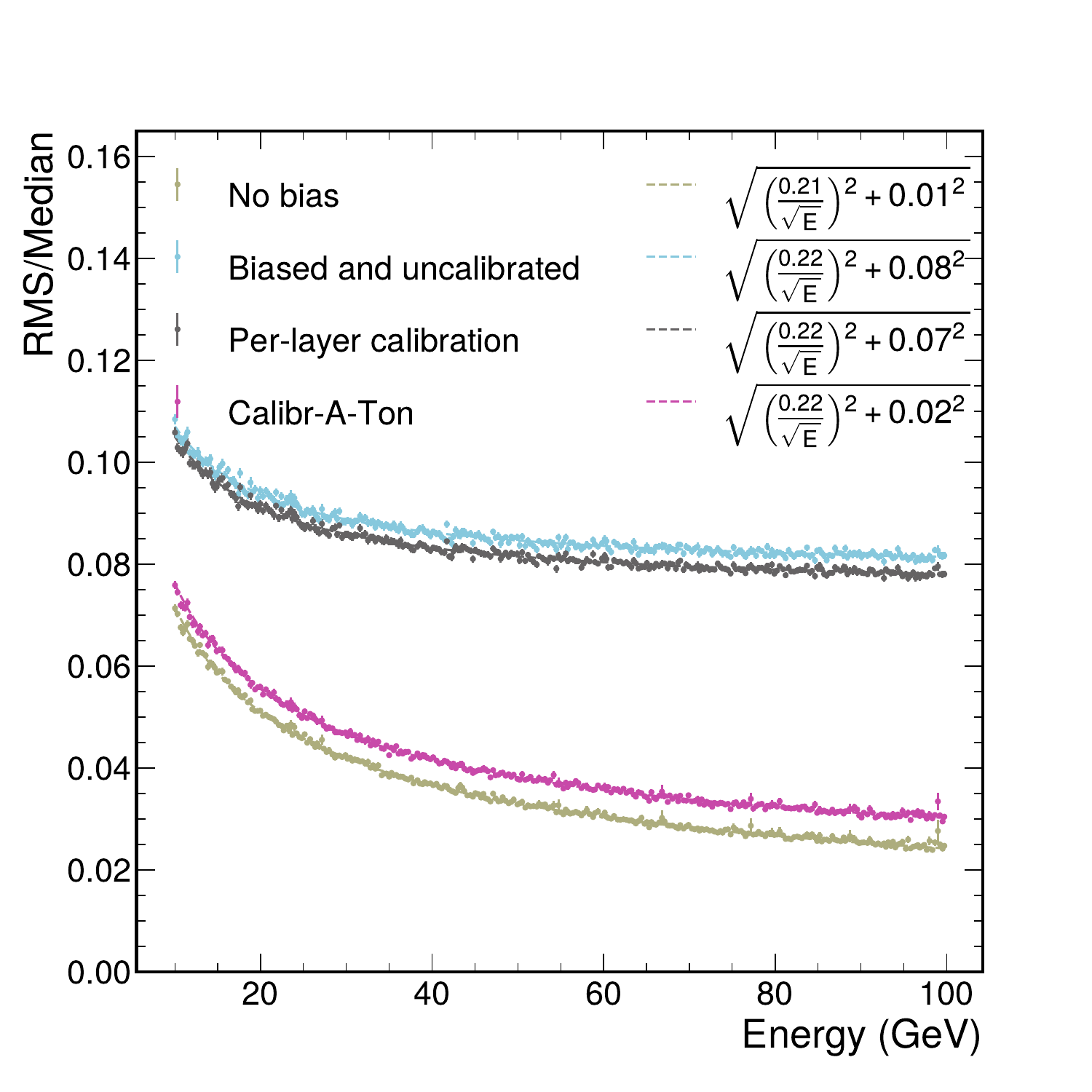}
\caption{Resolution on the shower energy (estimated by the root mean square RMS of the response histogram, divided by the median generator energy), as a function of the generator energy. Functions of the form $f(E_\mathrm{gen})=\sqrt{a^2/E_\mathrm{gen}+c^2}$ are fitted to these points. The resolution after application of the per-layer calibration constants (in black) still presents a sizable constant term ($c=0.07$). As illustrated by the magenta points, the calibration constants derived with the Calibr-A-Ton method are capable to recover the resolution observed in the unbiased case (dark green). Both cases present similar stochastic and constant terms ($a=0.22\;\sqrt{\mathrm{GeV}},c=0.02$ and $a=0.21\;\sqrt{\mathrm{GeV}},c=0.01$, respectively).}
\label{fig:resolution_after_training}
\end{center}
\end{figure}

The resolution, in bins of the electron generator energy, shown in Figure \ref{fig:resolution_after_training} for the different cases, confirms these conclusions. On the one hand, it can be observed that the per-layer calibration is unable to compensate for the biases introduced, and presents a resolution and a constant term ($c=0.07$) comparable to the uncalibrated, biased case ($c=0.08$). On the other hand, applying the calibration constants derived with the Calibr-A-Ton method leads to a resolution similar to the unbiased case, in the whole energy range. Similar stochastic and constant terms are extracted: $a=0.22\;\sqrt{\mathrm{GeV}}$, $c=0.02$ for Calibr-A-Ton when they were $a=0.21\;\sqrt{\mathrm{GeV}}$, $c=0.01$ in the unbiased case. Table~\ref{tab:resolution_parameters} summarizes these results. 
The achieved performance illustrates the potential of the Calibr-A-Ton method. It enables the derivation of calibration constants that depend on all relevant dimensions of the problem -- in this application, the 3D coordinates and energies of the cells. Additionally, it employs a custom binning approach that can be adapted to the available statistics along these dimensions. Finally, since each shower typically deposits energy in hundreds of cells, the method maximizes the statistical information available for determining numerous calibration constants. These features make Calibr-A-Ton a promising and highly suitable method for the calibration of high-granularity calorimeters such as the CMS HGCAL.

\begin{table}
\centering
\begin{tabular}{|l || c | c |} 
 \hline
  & $a$ $[\sqrt{\mathrm{GeV}}]$ & $c$ \\ [0.5ex] 
 \hline\hline
 Unbiased, uncalibrated & 0.21 & 0.01 \\\hline
 Biased, uncalibrated & 0.22 & 0.08 \\\hline
 Biased, per-layer calibration & 0.22 & 0.07 \\\hline
 Biased, Calibr-A-Ton & 0.22 & 0.02\\   
 \hline
\end{tabular}
\caption{Value of the resolution parameters for the stochastic term ($a$) and the constant term ($c$) in the different cases under study. Uncertainties on those parameters are negligible.}
\label{tab:resolution_parameters}

\end{table}

\section{Conclusion}
\label{conclusion}
This paper describes a novel method for the calibration of calorimeters, Calibr-A-Ton, that bypasses some of the major constraints found in historical calibration procedures. To test the implementation of the method, based on differential programming, and its performance, the simulation of a high granularity calorimeter, with one electromagnetic and two hadronic sections, is used. Simulated electrons of known energies are generated and produce shower in this detector, typically leaving energy in hundreds of cells. Known biases are then introduced on the cells energy response, to emulate the resolution effects that typically affect calorimeters. The results clearly show that, contrary to a simple approach based on layer-per-layer calibration, the Calibr-A-Ton method is able to meaningfully compensate for these biases and to nearly recover the original detector resolution. When fitting the resolution curve of the shower energy, almost identical stochastic and constant terms are extracted. On top of the performance achieved, it should be noted that Calibr-A-Ton is also more versatile than other methods, as it has the potential to be used in several contexts: simulation-based calibration (as is the case here), test beams,  and \emph{in situ} calibrations using physics signals or auxiliary measurements by another subdetector, such as the tracker.\\

This work represents a proof-of-principle of the Calibr-A-Ton method in the successful extraction of calibration constants for the electromagnetic section of a high-granularity calorimeter similar to the CMS HGCAL. Subsequent work will explore the calibration of the hadronic sections using simulated charged pion showers, the introduction of biases directly in the \texttt{GEANT4} simulation (material density fluctuations, presence of dead material, etc.), and the effects of a different choice of binning of the calibration constants on the performance.

\bibliographystyle{unsrt}
\bibliography{sample}

\end{document}